\documentclass{article}
\usepackage{setspace}
\doublespace
\usepackage[a4paper,left=3.3cm,right=3.3cm,top=3.4cm,bottom=3.4cm]{geometry} 

\usepackage[english]{babel} 			
\usepackage[utf8]{inputenc} 			

\usepackage{amsmath}					
\usepackage{bbm}						
\usepackage{amsfonts}
\usepackage{physics}					
\usepackage{mathtools}					

\usepackage[position = top]{floatrow}	

\usepackage{url}
\usepackage{natbib} 
\defcitealias{australian1997national}{ABS, 1997} 		
\defcitealias{abs2018houseocc}{ABS, 2019}
\defcitealias{abs2017hes}{ABS, 2017}

\usepackage{graphicx} 					
\graphicspath{{figures/}}				

\renewcommand{\arraystretch}{1.2}		

\begin{titlepage}
	\title{Incorporating Financial Hardship in Measuring the Mental Health Impact of Housing Stress}
	\author{Timothy Ludlow$^{1,2}$ \and Jonas Fooken$^{3}$ \and Christiern Rose$^{1}$ \and Kam Ki Tang$^{1}$}
	\date{\today}
	
	\footnotetext[1]{School of Economics, The University of Queensland}
	\footnotetext[2]{Corresponding author: timothy.ludlow1@uqconnect.edu.au}	
	\footnotetext[3]{Centre for the Business and Economics of Health, The University of Queensland}
	
\end{titlepage}

\begin{document}

\maketitle

\begin{abstract}

Housing expenditure tends to be sticky and costly to adjust, and makes up a large proportion of household expenditure. Additionally, the loss of housing can have catastrophic consequences. These specific features of housing expenditure imply that housing stress could cause negative mental health impacts. This research investigates the effects of housing stress on mental health, contributing to the literature by nesting housing stress within a measure of financial hardship, thus improving robustness to omitted variables and creating a natural comparison group for matching. Fixed effects (FE) regressions and a difference-in-differences (DID) methodology are estimated utilising data from the Household Income and Labour Dynamics in Australia (HILDA) Survey. The results show that renters who are in housing stress have a significant decline in self-reported mental health, with those in prior financial hardship being more severely affected. In contrast, there is little to no evidence of housing stress impacting on owners with a mortgage. The results also suggest that the mental health impact of housing stress is more important than some, but not all, aspects of financial hardship.  

\end{abstract}

\newpage

	\section{Introduction}\label{sec.intro}

A frequent research finding is that lower socioeconomic status is associated with higher rates of common mental health disorders, suggesting \textit{prima facie} that living and working in lower socioeconomic environments can lead to poor mental health outcomes \citep{dohrenwend1992socioeconomic, lorant2003socioeconomic}. A prominent explanation is that disadvantaged environmental circumstances can cause chronic arousal of stress pathways, with physiological disregulation increasing the risk of developing mental illness \citep{tafet2003psychoneuroendocrinological, fisher2010social}. Although a number of socioeconomic factors, such as poverty, low eduction, unemployment, and housing circumstances, are associated with poor mental health,  the highly correlated nature of socioeconomic variables makes causal pathways hard to identify \citep{fuchs2004reflections}. This research investigates the effects of housing stress on mental health. It contributes to the literature by nesting housing stress within a measure of financial hardship, allowing for a natural comparison group, and therefore increasing robustness to omitted variables and reverse causality.

Housing is a commonly investigated social determinant of health, with some researchers hypothesising that the experience of housing stress, where housing costs become overly burdensome on individual finances, can lead to declines in mental health, contributing towards the observed socioeconomic gradient in mental health \citep{bentley2011association, reeves2016reductions}. The economic features of housing expenditure means it has a particularly strong influence on personal finances, especially for those with limited resources. Housing costs make up a large proportion of total expenditure for most lower income individuals, and actions taken to reduce this expenditure can involve large adjustment costs. Additionally, housing is a necessity and, thus, the loss of housing can have catastrophic consequences. 

Previous research investigating the effect of housing stress on mental health has tended to find that unaffordable housing leads to small but significant declines in mental health \citep{mason2013housing, bentley2011association}. However the challenges of reverse causality and omitted variables suggest there are limitations to these results. Reverse causation is a concern as it is highly plausible that individuals with declining mental health are more prone to financial difficulties, including housing stress. This could occur, for example, if mental health declines are associated with more erratic or impulsive behaviour, which in turn increases the probability of experiencing housing stress. Omitted variables are a challenge for any observational setting and especially when investigating housing stress, as many socioeconomic variables are highly correlated \citep{fuchs2004reflections}. One concern is that the literature on housing stress tends to omit financial hardship from the analyses, despite financial hardship and mental health having strong associations \citep{fryers2003social, butterworth2009financial}. If results are based on a fixed effects (FE) regressions alone, the predominant methodology used in this research area, then they could be prone to bias in the presence of omitted variables and reverse causality.
 
A further complication encountered in this line of research is how housing stress should be measured. Although there is no general consensus on this issue, many researchers opt to use threshold based indicators to identify if an individual or household is under housing stress \citep{nepal2010measuring}. For example, a common approach is to define housing stress as a household or an individual spending more than 30 percent of their income on housing \citep{baker2020mental, mason2013housing, bentley2011association}. This definition may or may not be combined with another threshold that limits housing stress to households or individuals in the lower part of the income distribution, commonly the bottom 40 percent. These threshold type measures are useful policy variables for describing the prevalence of housing stress, and may be easily understood in public debate. However, using these definitions for empirical research at the individual level can be problematic. This is because small changes around the thresholds could have an undue influence on the results, especially when within variation alone is used to identify the effect of housing stress.
	If we consider housing stress to be a latent variable, missing housing payments are likely to be a strong signal of circumstances where housing costs are overly burdensome on personal finances. This seldomly utilised measure is a more direct signal of housing stress than threshold based financial indicators.
	
	This research classifies individuals that are unable to make rental or mortgage payments on time due to a shortage of money as being in housing stress. That is, housing stress is considered as a specific form of financial hardship. Accordingly, in the empirical analysis, we can differentiate between individuals experiencing other forms of financial hardship without housing stress, and those experiencing financial hardship with housing stress. As the former is more comparable to the latter than, say, those not experiencing any financial hardship, it provides a natural comparison group. This lends itself to a DID framework where those experiencing housing related and non-housing related financial hardship are the treatment group, and those experiencing non-housing related financial hardship only are the control group. Additionally, this research conducts FE regressions that account for both housing stress and financial hardship, a potential omitted variable in other papers.


This research is performed using longitudinal data from Australia, where modest government housing assistance payments, a limited supply of public housing options, and a relatively high exposure to private rental markets may increase the mental health impacts of housing stress \citep{bentley2016housing}. In Australia, the price-to-income ratio of housing has increased 78 percent between 1980 to 2015, and over the past two decades housing ownership rates have decreased from 70 percent to 66 percent of households \citep{hall2016housing}. Over the same period, the proportion of households living in public housing has reduced from just under 6 percent to around 3 percent, and the households renting in the private rental market has increased from 20 percent to 27 percent \citepalias{abs2018houseocc}. Housing is now the largest item of household expenditure, with lower-income households most affected.\footnote{Lower-income households are defined as the 38 percent between the $3^{rd}$ and $40^{th}$ percentiles of equivalised disposable households income.} For these lower income households, 57 percent of private renters and 41 percent of owners with a mortgage allocate more than 30 percent of their gross income to housing expenditure \citepalias{abs2018houseocc}.

Using longitudinal data from the Household Income and Labour Dynamics in Australia (HILDA) Survey, this research finds that renters in prolonged financial hardship are most impacted by housing stress. Using FE regressions, renters are found to have a $0.18$ standard deviation decline in self reported mental health when in housing stress, compared to having financial hardship without housing stress. The DID estimates reveal that renters in two previous periods of financial hardship that are exposed to housing stress have a 0.2 standard deviation decline in their mental health score, compared to similar individuals not exposed to housing stress. The results differ by tenure, with little or no effect found on mental health for mortgagors in housing stress, consistent with other research findings \citep{mason2013housing}. The results also suggest that the mental health impact of housing stress is more important than some, but not all, aspects of financial hardship. 


	\section{Economic Aspects of Housing Expenditure}\label{sec.prop.housing}

Housing makes up the largest proportion of household expenditure, which tends to be sticky and costly to adjust. These characteristics help explain why housing stress could impact on mental health. The ABS Household Expenditure Survey 2015-16 shows that housing makes up an average of 20 percent of household spending, more than any other category \citepalias{abs2017hes}.\footnote{Housing costs consist of rent, interest payments on mortgages, rates, home and content insurance, repairs and maintenance.} This proportion is even larger for households in the lower income or wealth quintiles. For individuals that enter into financial hardship, reducing the largest aspect of expenditure would be a logical course of action, however this can be complicated by high adjustment costs, making it hard to reduce housing expenditure at the margin. 

	The high adjustment costs associated with housing expenditure can be exacerbated by the nature of housing products as relatively indivisible and subject to long-term contracts. Housing is commonly subject to long-term contracts, either mortgages or leases, that lock in expenditure commitments into the future, either reducing the opportunities to move to cheaper alternatives, or increasing the cost of doing so. Additionally, housing is a relatively indivisible, or discrete product, that tends to be purchased as a whole by household units, further limiting expenditure adjustment opportunities, especially in the short-run. Sub-letting or sharing accommodation can reduce housing expenditures, however these too have their own adjustment and ongoing costs. 

	High adjustment costs can be seen as either direct or indirect. The direct costs are related to the actual event of moving. Besides the direct monetary costs, such as hiring a vehicle for moving furniture, there are also numerous time costs involved, such as searching for accommodation or packing possessions. These direct costs are higher for owners, who normally pay additional taxes, legal and agent fees. The indirect costs related to a move can be ongoing. For example, moving could require new employment with no guarantee of the same wage, or increased transportation costs if moving further away from a workplace. Welfare costs can also be high when moving if an individual has strong local connections to friends, family, educational institutions or other organisations in the community. 

Large housing expenditures combined with high adjustment costs are likely to make housing particularly salient for individuals in financial hardship. Additionally, the loss of housing can have catastrophic consequences. Not being able to meet housing expenditure commitments can be particularly serious when combined with other risk factors for homelessness, such as thin social support networks, limited earning capacity, exposure to domestic violence, or pre-existing mental or substance abuse disorders \citep{morrell2000becoming}. These features of housing expenditure imply that housing stress could have significant negative impacts on mental health.

	\section{Literature Review}\label{sec.lit.review}

Regression has been the most frequently employed technique to investigate the effect of housing stress on mental health. However, the potential for omitted variable bias and reverse causality lead to caution interpreting some results. \cite{reeves2016reductions} is an exception in the literature, utilising a quasi-experimental methodology that is able to address these concerns. The study uses a policy change in the U.K. that reduced the housing benefit paid to low income private renters, as an exogenous shock to investigate the impact of less affordable housing on depressive symptoms. The housing benefit expenditure is the second largest welfare payment in the U.K., and the 2011 policy change was a sizeable reduction of \pounds 1,220 per year.\footnote{This benefit reduction made up the largest individual contribution to budgetary savings amidst a range of austerity measures.}  Using a repeated cross section from the Annual Population Survey and a DID methodology, the effect of the policy change was a 1.8 percentage point increase in the probability of self-reported depressive symptoms by the benefit recipients, compared to individuals not receiving the benefit.\footnote{These results were robust to modelling choices, with similar results found using interrupted time-series analysis, and matching methods.} 

The more common approach in the literature is to combine longitudinal data with a FE methodology. This reduces omitted variable bias by removing time-invariant factors. \cite{bentley2011association} uses FE regressions to investigate the effect of unaffordable housing on mental health using data from the HILDA Survey. This finds that unaffordable housing causes small but significant decreases in self-reported mental health for individuals in the lower part of the income distribution when they transition into unaffordable housing.\footnote{Variables controlled for were age, equivalised disposable household income, and moving from one house to another.} This mental health decline was not seen for individuals in higher income households. \cite{mason2013housing} also uses a FE framework to investigate how the relationship between housing stress and mental health differs by tenure. Their results show that renters are more exposed than owners to the effects of unaffordable housing. \cite{bentley2016housing} uses FE regressions to investigate if the effect of housing stress on mental health differs based on the housing context across nations. Using longitudinal data from the U.K. and Australia, they find evidence that in the U.K. renters are offered some protection against the effects of unaffordable housing. The research of \cite{baker2020mental} uses both within and between variation to allow for inclusion of initial mental health status when considering the effect of prolonged exposure to unaffordable housing. Prolonged exposure is found to be significant, with the effect size larger in magnitude compared to intermittent exposure.

Besides the use of panel regression methods, \cite{bentley2011association}, \cite{mason2013housing} and \cite{baker2020mental} also employ a threshold approach to measuring housing stress, where individuals are considered to be in housing stress if they spend more than 30 percent of their income on housing.\footnote{Commonly gross income used, however there is no agreed income measure for this threshold.} This housing expenditure threshold is usually combined with an income threshold to reflect that spending a high proportion of income on housing is likely to create a greater burden for lower income earners. For example, \cite{bentley2011association} uses a 30 percent threshold for housing costs and tests for heterogeneity based on income thresholds. \cite{mason2013housing} also employs a 30 percent threshold, and only analyses individuals in the lower 40 percent of the income distribution. The approach used by \cite{mason2013housing} is known as the \textit{30/40 rule} definition of housing stress \citep{nepal2010measuring}. Although this rule is commonly used for policy discussion on housing affordability, especially in Australia, its arbitrary thresholds create limitations when used in research. This is especially the case when using FE models, where small variations around a threshold over time can make a disproportionately large contribution to parameter estimates. One solution is offered by \cite{taylor2007psychological}, which uses survey responses regarding problems with making mortgage or rental payments, and being more than two months behind on payments, as proxies for unsustainable housing commitments.



\cite{bentley2011association}, \cite{mason2013housing} and \cite{baker2020mental} all aim to isolate the effect of housing stress on mental health after controlling for household income, however do not control for financial hardship. 
	If financial hardship also causes mental health declines, and if it is positively correlated with housing stress, then it too should be included, or the effect of housing stress will be overstated.
	
	A separate stream of the literature has found that financial hardship is associated with mental health declines. 
	\cite{fryers2003social} conducts a review of this literature and finds that material circumstances are a marker of increased rates of common mental disorders. The researchers consider a number of indicators of common mental health disorders using a meta-analysis and find that material circumstances, along with education and employment, are the strongest markers. \cite{butterworth2009financial} looks directly at financial hardship indicators using the Personality and Total Health (PATH) Through Life Study. Using logistic regression of the second wave of the survey, and controlling for demographics, socioeconomic status, and prior depressive symptoms, the research finds a strong association between financial hardship and depression. \cite{kahn2006financial} also considers the effect of financial hardship on health, both physical and mental. Considering a sample of adults aged over 65 years and relying on recall of life experiences, they find that past financial hardship significantly determines a number of health outcomes, including depression, with persistence of hardship found to be particularly important. \cite{kiely2015financial} considers the effects of financial hardship on self-reported mental health, utilising the financial hardship indicators in the HILDA Survey. Its results from a series of logistic regressions are consistent with deprivation related hardship increasing the risk of mental health problems, with less predictive power associated with income poverty. 


The economics literature is well known for employing natural experiments and other causal approaches to address issues of endogeneity, and while there are contributions to investigating the effects of housing and location on mental health, housing stress has not specifically been addressed. \cite{cattaneo2009housing} provides evidence that housing quality improvements affect physical and mental health outcomes. The research analyses a government program in Mexico that replaced dirt with cement floors, taking advantage of variation in the timing of the program roll-out for identification. They find housing quality interventions improved the mental health of adults. In the Moving to Opportunity study, \cite{kling2007experimental} considers how neighbourhood effects can impact on mental health. Endogenous neighbourhood selection is overcome by using data from a randomised experiment where families in high-poverty U.S. housing projects are offered vouchers to move to lower-poverty neighbourhoods, with the results finding mental health benefits for adults, but mixed results for youth.


To summarise, research investigating the effects of housing stress on mental health commonly omits financial hardship and employs threshold based financial indicators to measure housing stress. This research focus on these potential limitations and employs commonly used econometric techniques to address robustness concerns. These techniques are outlined in the next section.




	\section{Identification}\label{sec.identification}

The following section outlines the identification strategy for estimating the effect of housing stress on mental health, compared to non-housing related financial hardship alone, with the key concept being a nesting of the housing stress measure within the financial hardship measures. Previous research commonly tries to find the effect of housing stress independent of financial circumstances, using income to control for financial circumstances (for example, \cite{bentley2011association}). However, in contrast to income, financial hardship indicators directly assess the impact of financial resources on life events, and have been found to be associated with mental health \citep{butterworth2009financial}. Excluding financial hardship from the mental health equation could upwardly bias the magnitude of the housing stress parameter, given financial hardship is positively correlated with housing stress, and negatively correlated with mental health.\footnote{This assumes that larger values indicate better mental health.} 

This research considers several types of financial hardship, with housing stress being one of them. Therefore, if an individual is in housing stress, by construction they must also be in financial hardship. This set-up is more robust against omitted variable bias, as it both accounts for financial hardship as an independent variable, and provides a natural comparison group, where individuals in housing stress are compared to those in financial hardship without housing stress.

The two models outlined below are estimated separately for owners with a mortgage and private renters, informed by previous research that has found differences based on tenure \citep{mason2013housing}. The baseline FE specification is first discussed, followed by the dynamic DID estimator.

		\subsection{Baseline Model} \label{sec.baseline.mod}

Consider the following equation that models mental health, $MH_{it}$, for each individual $i$ in time period $t$, as a function of housing stress: 
\begin{align}
MH_{it}= \alpha_i + \lambda_t + \tau \cdot \mbox{HousingStress}_{it} + X_{it}^{\prime} \beta +\epsilon_{it}\;,\label{eq.fe.estimation1}
\end{align}
where $\mbox{HousingStress}_{it}$ is an indicator equal to one if $i$ is in housing stress at time $t$; $\alpha_i$ is an individual specific FE; $\lambda_t$ is a time specific FE; $X_{it}$ is a vector of control variables that consists of equivalised disposable household income, the employment status (full-time, part-time, or otherwise\footnote{This category is defined as unemployed or not in the labour force.}), a measure of physical health (the SF-36 Physical Component Summary (PCS) lagged one period), and dummy variables indicating the quintile of disadvantage of the neighbourhood where the individual lives; $\epsilon_{it}$ is a random error term.

Equation (\ref{eq.fe.estimation1}) assumes that mental health is directly impacted by housing stress, with the impact of financial resources fully accounted for by income. However, as discussed earlier, if financial hardship has a direct impact on mental health after accounting for income, then $\tau$ could be bias. Financial hardship is therefore added to equation \ref{eq.fe.estimation1}, incorporating the nested structure of housing stress within financial hardship:
\begin{align}
MH_{it}= \alpha_i + \lambda_t + \theta \cdot \mbox{Hardship}_{it} +  \tau \cdot \mbox{HousingStress}_{it} + X_{it}^{\prime} \beta +\epsilon_{it}\;.\label{eq.fe.estimation2}
\end{align}
$\mbox{Hardship}_{it}$ is an indicator variable equal to one if $i$ is in financial hardship at $t$ and zero otherwise, and $\mbox{HousingStress}_{it}$ is now equivalent to the interaction term $(\mbox{Hardship}_{it} \times \mbox{HousingStress}_{it})$. These are equivalent because if $\mbox{HousingStress}_{it} = 1$, this implies that $\mbox{Hardship}_{it} = 1$.

Equation (\ref{eq.fe.estimation2}) is a two-way FE model that can be estimated from panel data with $N$ individuals and $T$ time periods using standard methods. The estimation of coefficients using within variation alone allows for individual specific, time-invariant omitted variables to be correlated with the dependent variables. The null hypothesis of $\tau = 0$ tests if housing stress has additional effects compared to non-housing related financial hardship alone, and can be tested with a t-test using cluster robust standard errors, clustering on the individual. For consistent estimation of the housing stress coefficient $\tau$, strict exogeneity is required. This could be seen as reasonable, as equation (\ref{eq.fe.estimation2}) accounts for financial hardship, a number of control variables, and time-invariant factors, although the assumption would not hold if there was reverse causality from mental health to housing stress. 

The uncertainty of the strict exogeneity assumption required for consistent FE estimates motivates a complementary DID methodology. This framework has the advantages of incorporating explicit matching based on treatment and pre-treatment periods, and allowing for the key identification assumption to be testable. Additionally, it allows us to explore heterogeneous effects from housing stress, based on prior financial hardship experience.		

		\subsection{Matched Difference-in-differences} \label{sec.het.causal.effects}

To assess the additional effect of housing stress on mental health, compared to non-housing related financial hardship alone, the ideal experiment would randomly assign individuals in non-housing related financial hardship to receive either housing stress (the treatment group) or no housing stress (the control group). Given data from treatment and control groups in the pre- and post-treatment periods, the average treatment effect (ATE) of housing stress, compared to non-housing related financial hardship alone, could be estimate using a number of methods, including a standard DID equation.\footnote{A number of methods are appropriate due to the random allocation of the treatment.}

To replicate this ideal experiment as closely as possible, this research proceeds by using a matching method. Individuals observed over three consecutive periods, with housing stress and non-housing related financial hardship in the third period, are considered as candidates for the treatment group. Similarly, individuals observed over three consecutive periods, with non-housing related financial hardship in the third period but no housing stress, are considered as candidates for the control group. The last of the three periods is referred to as the event (in reference to the event-studies literature), and an event time $t$ is constructed such that the event occurs at $t=0$. The two periods prior to the event, occurring at $t=-1$ and $t=-2$, are referred to as the pre-treatment time periods (consistent with DID terminology).

Based on financial hardship in the pre-treatment periods, candidate individuals are allocated into matched subgroups. Specifically, the \textit{Low} financial hardship group has no financial hardship in the pre-treatment periods, $t=-2$ and $t=-1$; the \textit{Moderate} financial hardship group has non-housing related financial hardship at $t=-1$, but not at $t=-2$; and the \textit{High} financial hardship group has two periods of non-housing related financial hardship in the pre-treatment periods, $t=-1$ and $t=-2$.

Within the same subgroup, some individuals have multiple candidates. When this is the case, only one set of three observations are kept based on two rules. First, if the individual is a candidate for both treatment and control groups, they are allocated to the treatment group only. Second, if the individual has multiple sets of three observations, only the first set of three observations (i.e. earliest based on calendar year) is selected.

This matching method is valid under the assumption that individuals in housing stress and non-housing related financial hardship are similar to individuals in non-housing related financial hardship alone, given the same prior financial hardship experience. If this assumption is valid, the control group can be used to form counterfactuals for the treatment group, and the additional effect of housing stress can be estimated. These effects can be heterogeneous based on prior financial hardship experience (i.e. based on the subgroup) and estimated using a dynamic DID estimator, with a test for parallel trends equivalent to testing the validity of the matching procedure.



More formally, for individual $i$ in (time invariant) group $g \in \{\mbox{Treat}, \mbox{Control}\}$, at event time $t$, unconfoundedness is assumed in the following form:
\begin{align*}
\mathbb{E}[MH_{igt}(0)| \alpha_g, \boldsymbol{D_{g}}, \boldsymbol{X_{i}}, \mbox{Hardship}^{\prime} = h] = \mathbb{E}[MH_{igt}(0)| \alpha_g, \boldsymbol{X_{i}}, \mbox{Hardship}^{\prime} = h]\\
\mathbb{E}[MH_{igt}(1)| \alpha_g, \boldsymbol{D_{g}}, \boldsymbol{X_{i}}, \mbox{Hardship}^{\prime} = h] = \mathbb{E}[MH_{igt}(1)| \alpha_g, \boldsymbol{X_{i}}, \mbox{Hardship}^{\prime} = h]
\end{align*}
where $MH_{igt}^{(h)}(0)$ and $MH_{igt}^{(h)}(1)$ are the potential outcomes under the control and treatment event respectively; $\boldsymbol{D_{g}}$ represents the history of housing stress such that $\boldsymbol{D_{g}} = \{D_{g,-2}, D_{g,-1}, D_{g,0}\}$, where $D_{g,t}$ is equal to one if individual $i$ with $g=\mbox{Treat}$ is in $t=0$, and zero otherwise (this amounts to $D_{Treat}=\{0,0,1\}$ and $D_{Control}=\{0,0,0\}$); the vector $\boldsymbol{X_{i}}$ represents the history of control variables such that $\boldsymbol{X_{i}} = \{X_{i,-2}, X_{i,-1}, X_{i,0}\}$, and $\alpha_g$ is a group specific FE that captures time-invariant differences between treatment and control groups. The variable $\mbox{Hardship}^{\prime}$ represents the history of financial hardship that the entire subsample is conditioned on, such that $h \in \{\{0,0,1\}, \{0,1,1\}, \{1,1,1\}\}$. The remaining notation suppresses the conditioning on $h$, instead using a superscript $(h)$ to indicate dependence on the subsample.

The treatment effect is assumed to be additive such that 
\begin{align*}
\mathbb{E}[MH_{igt}^{(h)}(1)| \alpha_g, \boldsymbol{X_{i}}] = \mathbb{E}[MH_{igt}^{(h)}(0)| \alpha_g, \boldsymbol{X_{i}}] + \tau_{t=0}^{(h)}
\end{align*}
where $\tau_{t=0}^{(h)}$ is the heterogeneous treatment effect equal to a constant $\tau^{(h)}$ at event time $t=0$, and zero otherwise.\footnote{In the additive treatment effect equation, conditioning on treatment status is not required because unconfoundedness is assumed.}

Fundamental to the potential outcomes framework is that both potential outcomes cannot be observed, only $MH^{(h)}_{igt}=(1-D_{gt}) \times MH^{(h)}_{igt}(0) + D_{gt} \times MH^{(h)}_{igt}(1)$. Then
\begin{align*}
\mathbb{E}[MH_{igt}^{(h)}| \alpha_g, \boldsymbol{D_{g}}, \boldsymbol{X_{i}}] = \mathbb{E}[MH_{igt}^{(h)}(0)| \alpha_g, \boldsymbol{X_{i}}] + \tau^{(h)}D_{igt}.
\end{align*}
Making the following functional form assumption
\begin{align*}
\mathbb{E}[MH_{igt}^{(h)}(0)| \alpha_g, \boldsymbol{X_{i}}] = \alpha_{g}^{(h)} + \lambda_t^{(h)} + X_{it}^{\prime} \beta^{(h)},
\end{align*}
the conditional expectations function can be written as
\begin{align}
\label{eq.derive.did3}
\mathbb{E}[MH_{igt}^{(h)}| \alpha_g, \boldsymbol{D_{g}}, \boldsymbol{X_{i}}] = \alpha_{g}^{(h)} + \lambda_t^{(h)} + X_{it}^{\prime} \beta^{(h)} + \tau^{(h)}D_{igt} \,.
\end{align}

Equation (\ref{eq.derive.did3}) assumes a common trend is followed by all individuals, represented by $\lambda^{(h)}_t$. This follows directly from the unconfoundedness assumption and is equivalent to the parallel trends assumption in a DID framework. Equation (\ref{eq.derive.did3}) can be modified to include separate trends for the treatment and control groups to assess parallel trends via visual inspection and hypothesis testing. Thus, estimation of equation (\ref{eq.derive.did3}) is given by the dynamic DID model
\begin{align}
\label{eq.did.estimation}
\begin{split}
MH_{it}^{(h)}= \,&\alpha_1^{(h)} \,+ \alpha_2^{(h)} \cdot \mbox{Treat}_{i}\, +\, \gamma_1^{(h)} \cdot \mathbbm{1}[t=-1]\, +\, \gamma_2^{(h)} \cdot \mathbbm{1}[t=-1] \cdot \mbox{Treat}_{i} \\ 
&+\, \delta^{(h)} \cdot \mathbbm{1}[t=0]\, +\, \tau^{(h)} \cdot \mathbbm{1}[t=0] \cdot \mbox{Treat}_{i}\, +\, X^{\prime}_{it} \beta^{(h)} \, + \, \epsilon_{it}\;,
\end{split}
\end{align}
where $\mbox{Treat}_i$ is an indicator variable equal to one if individual $i$ is in the treatment group, and zero if in the control group; and $\mathbbm{1}[t=s], \,s\in\{-2,-1\}$ is a time dummy equal to one when $t=s$ and zero otherwise.  To account for any time effects from the calendar year when the event occurs, event time FEs are added as additional control variables to $X_{it}$.

As all individuals in the control group experience financial hardship at $t=0$, this effect is fully captured by $\delta^{(h)}$. The parameter $\tau^{(h)}$ represents the additional effect of housing stress, compared to individuals that have only non-housing related financial hardship. Equation (\ref{eq.did.estimation}) can be estimated using least squares with cluster robust standard errors, clustering on the individual.

The parallel trends assumption can be inspected visually, and tested explicitly with a t-test under the null hypothesis $\gamma_{2}^{(h)}=0$; evidence of a non-zero coefficient corresponds to a rejection of the parallel trends assumption. 

	\section{Data}\label{sec.data}

The longitudinal data used for the analysis is from the HILDA Survey, an ongoing representative survey of Australian households that has been conducted annually since 2001. The survey collects detailed information based around the broad areas of income, labour and family dynamics, and includes information on health status and housing. Table \ref{tab.ss.full0119} provides summary statistics for the variables used in this research. The summary statistics are for the years 2001 and 2019, the first and last cross sections of the panel, and includes all responding adults in the sample.\footnote{To maintain the representativeness of the survey, a top-up sample was added in wave 11, accounting for the increased number of observation in 2019 compared to 2001. Adults are considered aged 15 years and over. A responding adult is someone from a responding household that completes an interview.}

From Table \ref{tab.ss.full0119}, we see that in 2001 around 72 percent of sample respondents had ownership of their place of residence (with or without a mortgage), while 26 percent were renters. Ownership rates in 2019 are lower, down to 66 percent, while the proportion of renters increases to 31 percent. Furthermore, the increasing proportion of renters is associated with increased participation in the private rental market (that is, renting from a private landlord), increasing from 20 percent in 2001 to 27 percent in 2019, while those renting from a public housing authority decreased from 4 percent to 3 percent over the same period. These patterns are consistent with that reported from Survey of Income and Housing (SIH) conducted by the Australian Bureau of Statistics \citepalias{abs2018houseocc}.\footnote{For example, the SIH shows, from 1997-8 to 2017-18, private renters increased from 20 percent to 27 percent, and ownership fell from 70 percent to 66 percent, as discussed in Section \ref{sec.intro}.} 

The two mental health measures used as dependent variables for the analysis are shown in Table \ref{tab.ss.full0119}: These are the Mental Component Summary and the Mental Health Scale, both of which are derived from the Short Form-36 (SF-36) Survey. The SF-36 is an internationally recognised tool for assessing an individual's functional health status and well-being. It has been found to produce valid and reliable results at both clinical and population levels, and is collected in each wave of the HILDA Survey \citepalias{australian1997national}. Aggregating responses from five of the thirty-six questions related to mental health provides one of eight aggregate scales, the Mental Health Scale\footnote{The Mental Health Scale is transformed to range from 0 - 100.}. These eight aggregate scales can be further summarised into two measures representing the physical and mental dimensions of health, referred to as the Physical Component Summary (PCS) and the Mental Component Summary (MCS). The MCS and PCS range from 0-100, have a standard deviation of 10, and higher values represent better health (see appendix \ref{sec.mcs.construct} for further details). Due to the common usage, validity and reliability of the MCS, it is chosen as the primary measure for mental health, with the Mental Health Scale used for robustness checks.

\begin{table}[!htbp] \centering \renewcommand*{\arraystretch}{1.1}\caption{Full Sample Summary Statistics}\label{tab.ss.full0119}\resizebox{\textwidth}{!}{
\begin{tabular}{lrrrrrr}
\hline
\hline
\textit{Year} & \multicolumn{3}{c}{2001} & \multicolumn{3}{c}{2019}  \\ 
\hline
\textit{Variable} & \multicolumn{1}{c}{N} & \multicolumn{1}{c}{Mean} & \multicolumn{1}{c}{SD} & \multicolumn{1}{c}{N} & \multicolumn{1}{c}{Mean} & \multicolumn{1}{c}{SD} \\ 
\hline
Owner & 13969 & 0.72 & 0.45 & 17441 & 0.66 & 0.47 \\ 
Owner with mortgage & 13969 & 0.31 & 0.46 & 17462 & 0.36 & 0.48 \\ 
Monthly mortgage payments & 4394 & 1503 & 1160 & 6204 & 2096 & 1750 \\ 
Renter & 13969 & 0.26 & 0.44 & 17441 & 0.31 & 0.46 \\ 
Rents from private landlord & 13969 & 0.2 & 0.4 & 17455 & 0.27 & 0.45 \\ 
Rents from public housing & 13969 & 0.04 & 0.2 & 17455 & 0.03 & 0.18 \\ 
Monthly rental payments & 3573 & 1051 & 565 & 5495 & 1490 & 838 \\ 
\hline
Financial hardship & 12484 & 0.35 & 0.48 & 15530 & 0.26 & 0.44 \\ 
Missed housing payments & 12718 & 0.09 & 0.29 & 15707 & 0.06 & 0.23 \\ 
Could not pay bills & 12848 & 0.19 & 0.39 & 15765 & 0.11 & 0.31 \\ 
Sold or pawned something & 12772 & 0.07 & 0.25 & 15733 & 0.06 & 0.24 \\ 
Went without meals & 12783 & 0.05 & 0.21 & 15744 & 0.05 & 0.21 \\ 
Went without heating & 12762 & 0.04 & 0.19 & 15732 & 0.03 & 0.18 \\ 
Sort help from friends or family & 12816 & 0.17 & 0.37 & 15767 & 0.13 & 0.33 \\ 
Sort help from welfare or community organisation & 12787 & 0.05 & 0.23 & 15732 & 0.04 & 0.2 \\ 
Unable to raise \$2000 (\$3000) for emergency & 12849 & 0.17 & 0.37 & 15832 & 0.12 & 0.33 \\ 
\hline
Mental component summary & 12323 & 48.59 & 10.4 & 15698 & 47.25 & 11.32 \\ 
Mental health scale & 12933 & 73.72 & 17.48 & 15987 & 72.18 & 18.23 \\ 
Physical component summary & 12323 & 49.68 & 10.5 & 15698 & 49.16 & 10.69 \\ 
\hline
Age & 13969 & 43.35 & 17.7 & 17462 & 45.78 & 19.27 \\ 
Male & 13969 & 0.53 & 0.5 & 17462 & 0.53 & 0.5 \\ 
Employed full-time & 13969 & 0.42 & 0.49 & 17462 & 0.42 & 0.49 \\ 
Employed part-time & 13969 & 0.2 & 0.4 & 17462 & 0.21 & 0.41 \\ 
Neighbourhood quintile of disadvantage & 13969 & 2.94 & 1.44 & 17448 & 3.04 & 1.41 \\ 
Eqv. household disposable income (\$1,000's) & 13969 & 42.44 & 29.82 & 17462 & 57.61 & 39.58\\ 
\hline
\hline
\end{tabular}
}
\end{table}



		\subsection{Sample Data Based on Tenure}
		
Analysis of the HILDA Survey data is performed based on tenure. Specifically, the data is divided into owners with a mortgage, and renters in the private rental market. Table \ref{tab.ss.tenure19} shows the summary statistics by tenure for the year 2019. The mean of the two mental health variables are both lower for the renters compared to owners, with the a mean MCS value of 47.5 for owners, and 44.6 for renters. On average, owners have higher housing costs than renters with a mean monthly mortgage payment of \$2096, compared to monthly rental payments of \$1581.  Owners on average have higher incomes, live in less disadvantaged neighbourhoods, are older, and have slightly lower physical health. 


	\subsubsection{Housing Stress and Financial Hardship Variables}

Table \ref{tab.ss.tenure19} shows renters are more likely to be in financial hardship compared to owners, with 44 percent in hardship compared to 22 percent of owners. This measure of financial hardship is based on the response to a financial hardship survey question in HILDA that assesses if individuals are short of money, combined with a further question assessing their ability to raise emergency funds. 

Specifically, survey participants are first asked \textit{Since January did any of the following happen to you because of a shortage of money?} The seven options survey participants can select are: \textit{Could not pay electricity, gas or telephone bills on time; Could no pay the mortgage or rent on time; Pawned or sold something; Went without meals; Was unable to heat home; Asked for financial help from friends or family;} and \textit{Asked for help from welfare / community organisations}. Multiple options can be chosen from this first question. The second survey question is \textit{Suppose you had only one week to raise \$2000 (\$3000 waves nine onwards) for an emergency. Which of the following best describes how hard it would be for you to get that money?} Survey participants can select one of four responses: \textit{I could easily raise the money; I could raise the money, but it would involve some sacrifices; I would have to do something drastic to raise the money; I don't think I could raise the money.}

If an individual indicates that any of the seven events in the first question have occurred, or responds \textit{I don't think I could raise the money} to the second question, they are classified as being in financial hardship.\footnote{Whereas the first question directly asks about financial hardship events, the the second question is an indication of financial insecurity.} Individuals responding that they are unable to pay their rent or mortgage on time due to a shortage of money are classified as being in housing stress, as well as in financial hardship. Individuals that are not in housing stress, but belonging to any of the seven other hardship categories, are referred to as being in non-housing related financial hardship. 
 
The eight financial hardship indicators used for this research, derived from the two questions discussed above, are identical to the \textit{financial stress experiences} collected in the ABS Household Expenditure Survey (HES), except for the HES collects one extra category regarding non-payment of registration or insurance.\footnote{The ABS defines financial stress differently to how financial hardship is defined here, as they combine the questions regarding \textit{financial stress experiences} with \textit{missing out experiences}. These \textit{missing out experiences} are not available in the HIDLA Survey.}

\begin{figure}
\centering
\caption{Financial Hardship Categories by Tenure, 2019}
\includegraphics[scale=.62]{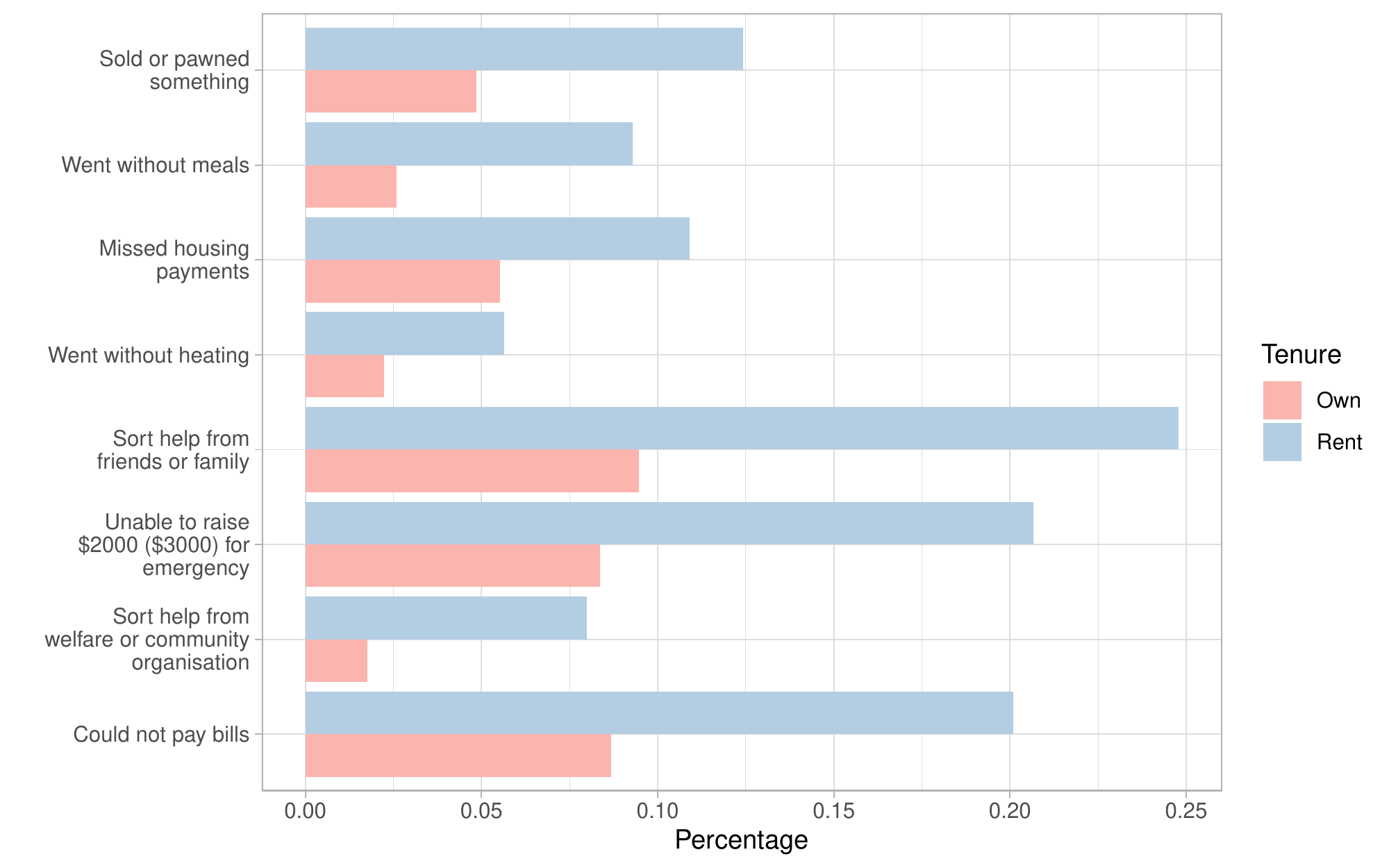}
\label{fig.ss.finhsvars}
\floatfoot{\textit{Notes:} Exact questionnaire wording: \textit{Since January did any of the following happen to you because of a shortage of money?} Response options: \textit{Could not pay electricity, gas or telephone bills on time; Could no pay the mortgage or rent on time; Pawned or sold something; Went without meals; Was unable to heat home; Asked for financial help from friends or family;} and \textit{Asked for help from welfare / community organisations}. Additional question: \textit{Suppose you had only one week to raise \$2000 (\$3000 waves nine onwards) for an emergency. Which of the following best describes how hard it would be for you to get that money?} Response: \textit{I don't think I could raise the money.}}
\end{figure}

Figure \ref{fig.ss.finhsvars} shows the proportion of renters and owners experiencing each of the eight financial hardship categories for the year 2019. For both owners and renters, the most commonly experienced category of financial hardship is asking for financial help from friends or family; 25 percent of private renters and 9 percent of owners experienced this hardship. The inability to pay electricity, gas or telephone bills on time, or to raise money in the case of an emergency, are also frequently experienced categories. Around 11 percent of renters in hardship indicate they have missed rental payments due to lack of money, while 6 percent of owners miss mortgage repayments for the same reason.

\begin{table}[!htbp] \centering \renewcommand*{\arraystretch}{1.1}\caption{Summary Statistics By Tenure, 2019}\label{tab.ss.tenure19}\resizebox{\textwidth}{!}{
\begin{tabular}{lrrrrrr}
\hline
\hline
\textit{Tenure} & \multicolumn{3}{c}{Owners with Mortgage} & \multicolumn{3}{c}{Private Renters}  \\ 
\hline
\textit{Variable} & \multicolumn{1}{c}{N} & \multicolumn{1}{c}{Mean} & \multicolumn{1}{c}{SD} & \multicolumn{1}{c}{N} & \multicolumn{1}{c}{Mean} & \multicolumn{1}{c}{SD} \\ 
\hline
Monthly mortgage payments & 6204 & 2097 & 1750 & - & - & -  \\ 
Monthly rental payments & - & - & - & 4759 & 1581 & 772 \\ 
\hline
Financial hardship & 5572 & 0.22 & 0.41 & 4052 & 0.44 & 0.5 \\ 
Missed housing payments & 5622 & 0.06 & 0.23 & 4119 & 0.11 & 0.31 \\ 
Could not pay bills & 5629 & 0.09 & 0.28 & 4137 & 0.2 & 0.4 \\ 
Sold or pawned something & 5621 & 0.05 & 0.21 & 4124 & 0.12 & 0.33 \\ 
Went without meals & 5622 & 0.03 & 0.16 & 4125 & 0.09 & 0.29 \\ 
Went without heating & 5622 & 0.02 & 0.15 & 4118 & 0.06 & 0.23 \\ 
Sort help from friends or family & 5628 & 0.09 & 0.29 & 4136 & 0.25 & 0.43 \\ 
Sort help from welfare or community organisation & 5618 & 0.02 & 0.13 & 4122 & 0.08 & 0.27 \\ 
Unable to raise \$2000 (\$3000) for emergency & 5662 & 0.08 & 0.28 & 4162 & 0.21 & 0.4 \\ 
\hline
Mental component summary & 5630 & 47.45 & 10.58 & 4166 & 44.61 & 12.14 \\ 
Mental health scale & 5704 & 72.87 & 17.07 & 4229 & 68.21 & 19.26 \\ 
Physical component summary & 5630 & 51.67 & 8.77 & 4166 & 50.56 & 10.43 \\ 
\hline
Age & 6204 & 40.48 & 14.49 & 4762 & 36.6 & 15.63 \\ 
Male & 6204 & 0.51 & 0.5 & 4762 & 0.52 & 0.5 \\ 
Employed full-time & 6204 & 0.58 & 0.49 & 4762 & 0.49 & 0.5 \\ 
Employed part-time & 6204 & 0.25 & 0.43 & 4762 & 0.21 & 0.41 \\ 
Neighbourhood quintile of disadvantage & 6203 & 3.2 & 1.36 & 4761 & 2.88 & 1.41 \\ 
Eqv. household disposable income (\$1,000's) & 6204 & 67.86 & 40.13 & 4762 & 50.41 & 30.7\\ 
\hline
\hline
\end{tabular}
}
\end{table}


\begin{figure}
\centering
\caption{Severity of Financial Hardship, 2019}
\includegraphics[scale=.6]{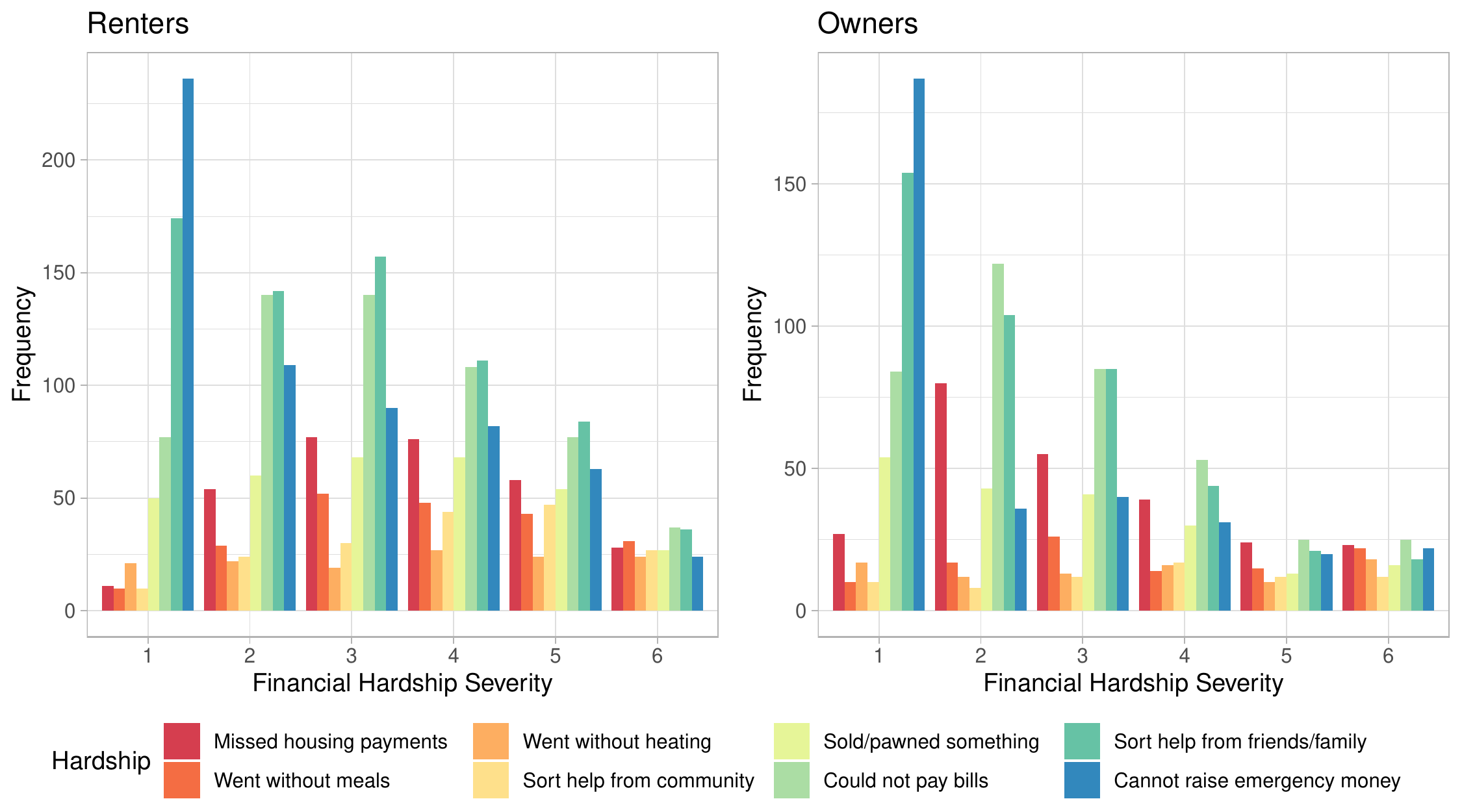}
\label{fig.sum.finhs.count}
\floatfoot{\textit{Notes:} Uses individuals in financial hardship, 2019. For individuals with a given level of financial hardship severity, displays the frequency of each financial hardship category. The sum of all eight financial hardship indicator variables is used as a proxy for the severity of financial hardship (thus, eight represents maximum hardship). The two most severe hardship categories are relatively uninformative and not displayed. \textit{Left:} Private renters. \textit{Right:} Owners with a mortgage.}
\end{figure}	


It is informative to consider how each of the financial hardship categories are associated with the severity of financial hardship. For individuals experiencing financial hardship in 2019, Figure \ref{fig.sum.finhs.count} uses the total number of financial hardship categories an individual is in as a proxy for severity, and shows the frequency of each category by severity. For individuals in \textit{mild} hardship (experiencing only one financial hardship category), borrowing money from friends or family and not being able to raise emergency funds are very frequent occurrences. It is also rare for renters or owners in \textit{mild} hardship to miss housing payments. However, the frequency of missed housing payments increases substantially for individuals in \textit{moderate} hardship (experiencing multiple financial hardship categories).

		\subsection{Matched Subsamples}
		
\begin{figure}
\centering
\caption{Time of Event - Matched Subsamples}
\includegraphics[scale=.6]{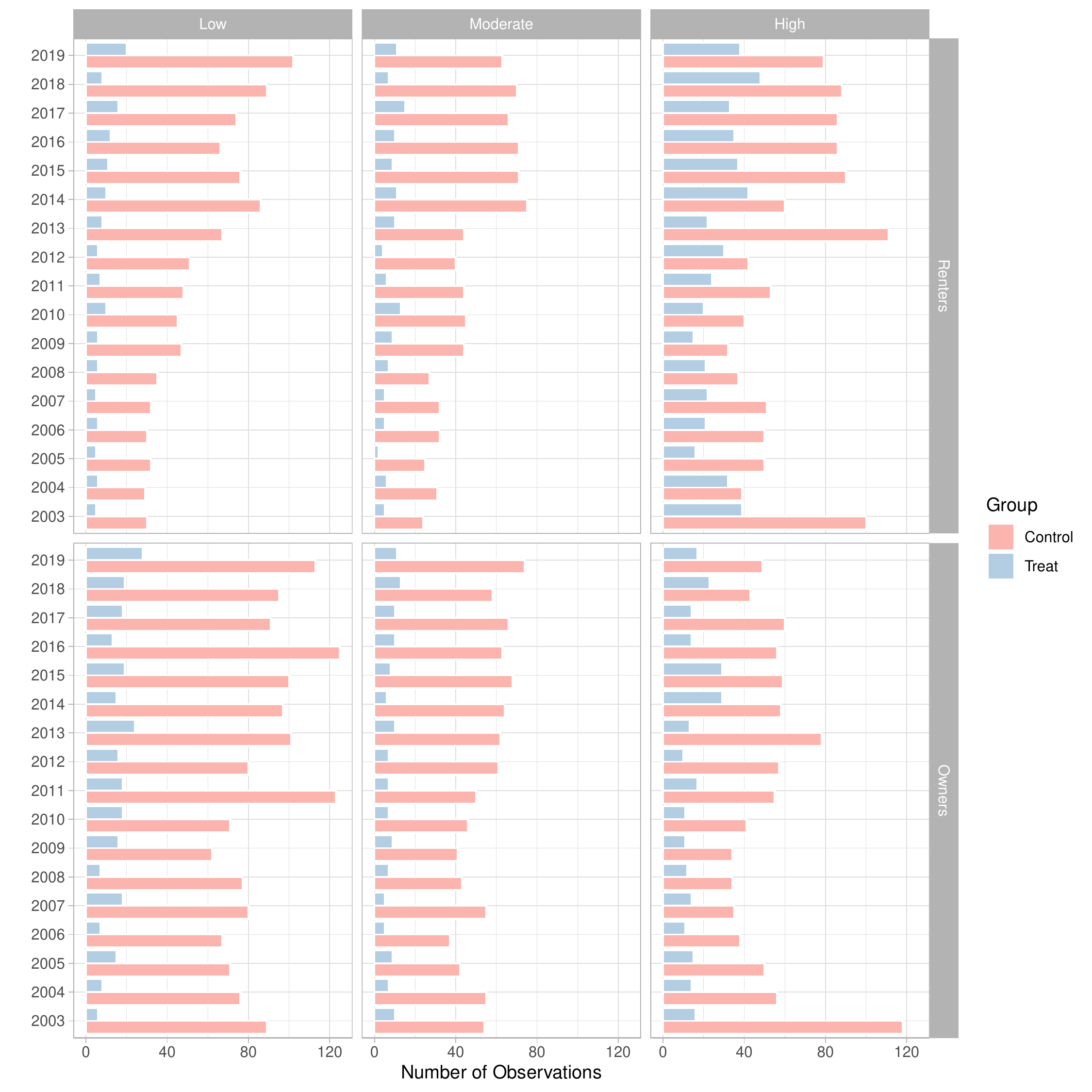}
\label{fig.ss.subgroups.events}
\floatfoot{\textit{Notes:} The treatment group has an event of housing stress. The control group has an event of non-housing related financial hardship. The \textit{Low} subgroup requires two preceding periods of no financial stress; the \textit{Moderate} subgroup has no financial hardship at $t=-2$, and non-housing related financial hardship at $t=-1$; the \textit{High} subgroup has two per-treatment periods of non-housing related financial hardship.}
\end{figure}

As detailed in section \ref{sec.het.causal.effects}, matched subsamples are formed for the DID analysis, resulting in \textit{Low}, \textit{Moderate}, and \textit{High} subsamples for both owners and renters. Each sample consists of treatment and control group observations, where the treatment group has an event of housing stress at $t=0$, and the control group has an event of non-housing related financial hardship. Figure \ref{fig.ss.subgroups.events} shows the calendar year these events occur for each matched subsample, illustrating events occur evenly across calendar years.



			\subsubsection{Renters and Owners Matched Subsamples}\label{sec.matchss}

Tables \ref{tab.ss.subgroups.low}, \ref{tab.ss.subgroups.mod}, and \ref{tab.ss.subgroups.high} show the matched subsamples for renters and owners at event time $t=-1$, split by treatment status. There are differences in the number of individuals in each subgroup, with as few as 125 individuals in the \textit{Moderate} treatment group for renters, and as many as 1460 in the \textit{Low} control group for owners. It is worth noting that it is the number of observations in the treatment group of each subsample that constrains the number of pre-treatment periods to two.\footnote{The methodology could easily be adapted to include more pre-treatment periods, given a larger sample size.}

The matching methodology is seen to perform well when we compare the averages of the matched treatment and control observations. For example, the treatment and control observations of the \textit{Low} subgroup for renters have similar mean values for mental health (MCS), disposable income, physical health (PCS), quintile of neighbourhood disadvantage, and employment. Between the subgroups we do see differences. As expected, the presence of non-housing related financial hardship in the pre-treatment periods is associated with decreases in the average MCS, along with disposable income and other variables.

\begin{table}[!htbp] \centering \renewcommand*{\arraystretch}{1.1}\caption{Summary Statisics - Low Matched Subgroups (t = -1)}\label{tab.ss.subgroups.low}\resizebox{\textwidth}{!}{
\begin{tabular}{l|rrrrrr|rrrrrr|}
\hline
\hline
\textit{Subgroup} & \multicolumn{3}{c}{Renters: Control} & \multicolumn{3}{c|}{Renters: Treat} & \multicolumn{3}{c}{Owners: Control} & \multicolumn{3}{c|}{Owners: Treat}  \\ 
\textit{Variable} & \multicolumn{1}{c}{N} & \multicolumn{1}{c}{Mean} & \multicolumn{1}{c}{SD} & \multicolumn{1}{c}{N} & \multicolumn{1}{c}{Mean} & \multicolumn{1}{c|}{SD} & \multicolumn{1}{c}{N} & \multicolumn{1}{c}{Mean} & \multicolumn{1}{c}{SD} & \multicolumn{1}{c}{N} & \multicolumn{1}{c}{Mean} & \multicolumn{1}{c|}{SD} \\ 
\hline
Mental component summary & 913 & 47.33 & 10.46 & 141 & 47.24 & 11.27 & 1460 & 48.12 & 9.85 & 261 & 48.01 & 10.3 \\ 
Employed full-time & 913 & 0.51 & 0.5 & 141 & 0.57 & 0.5 & 1460 & 0.56 & 0.5 & 261 & 0.62 & 0.49 \\ 
Employed part-time & 913 & 0.22 & 0.42 & 141 & 0.23 & 0.42 & 1460 & 0.27 & 0.44 & 261 & 0.2 & 0.4 \\ 
Neighbourhood disadvantage quintile & 913 & 3 & 1.39 & 141 & 2.94 & 1.54 & 1460 & 3.1 & 1.36 & 261 & 3 & 1.41 \\ 
Disposable income (\$1,000's) & 913 & 47.21 & 22.14 & 141 & 49.4 & 23.82 & 1460 & 56.26 & 26.14 & 261 & 58.13 & 36.47 \\ 
Physical component summary & 913 & 51.37 & 9.21 & 141 & 52.05 & 8.47 & 1460 & 51.48 & 8.68 & 261 & 51.03 & 8.43\\ 
\hline
\hline
\end{tabular}
}
\end{table}

\begin{table}[!htbp] \centering \renewcommand*{\arraystretch}{1.1}\caption{Summary Statisics - Moderate Matched Subgroups (t = -1)}\label{tab.ss.subgroups.mod}\resizebox{\textwidth}{!}{
\begin{tabular}{l|rrrrrr|rrrrrr|}
\hline
\hline
\textit{Subgroup} & \multicolumn{3}{c}{Renters: Control} & \multicolumn{3}{c|}{Renters: Treat} & \multicolumn{3}{c}{Owners: Control} & \multicolumn{3}{c|}{Owners: Treat}  \\ 
\textit{Variable} & \multicolumn{1}{c}{N} & \multicolumn{1}{c}{Mean} & \multicolumn{1}{c}{SD} & \multicolumn{1}{c}{N} & \multicolumn{1}{c}{Mean} & \multicolumn{1}{c|}{SD} & \multicolumn{1}{c}{N} & \multicolumn{1}{c}{Mean} & \multicolumn{1}{c}{SD} & \multicolumn{1}{c}{N} & \multicolumn{1}{c}{Mean} & \multicolumn{1}{c|}{SD} \\ 
\hline
Mental component summary & 770 & 45.5 & 11.58 & 125 & 45.67 & 11.36 & 898 & 46.46 & 10.67 & 139 & 44.92 & 12.26 \\ 
Employed full-time & 770 & 0.39 & 0.49 & 125 & 0.45 & 0.5 & 898 & 0.47 & 0.5 & 139 & 0.55 & 0.5 \\ 
Employed part-time & 770 & 0.25 & 0.44 & 125 & 0.28 & 0.45 & 898 & 0.3 & 0.46 & 139 & 0.24 & 0.43 \\ 
Neighbourhood disadvantage quintile & 770 & 2.77 & 1.39 & 125 & 2.61 & 1.34 & 898 & 2.91 & 1.34 & 139 & 2.81 & 1.41 \\ 
Disposable income (\$1,000's) & 770 & 41.86 & 18.74 & 125 & 43.32 & 17.69 & 898 & 50.06 & 24.38 & 139 & 49.73 & 22.57 \\ 
Physical component summary & 770 & 50.48 & 10 & 125 & 51.95 & 8.66 & 898 & 51.03 & 9.3 & 139 & 50.75 & 9.58\\ 
\hline
\hline
\end{tabular}
}
\end{table}

\begin{table}[!htbp] \centering \renewcommand*{\arraystretch}{1.1}\caption{Summary Statisics - High Matched Subgroups (t = -1)}\label{tab.ss.subgroups.high}\resizebox{\textwidth}{!}{
\begin{tabular}{l|rrrrrr|rrrrrr|}
\hline
\hline
\textit{Subgroup} & \multicolumn{3}{c}{Renters: Control} & \multicolumn{3}{c|}{Renters: Treat} & \multicolumn{3}{c}{Owners: Control} & \multicolumn{3}{c|}{Owners: Treat}  \\ 
\textit{Variable} & \multicolumn{1}{c}{N} & \multicolumn{1}{c}{Mean} & \multicolumn{1}{c}{SD} & \multicolumn{1}{c}{N} & \multicolumn{1}{c}{Mean} & \multicolumn{1}{c|}{SD} & \multicolumn{1}{c}{N} & \multicolumn{1}{c}{Mean} & \multicolumn{1}{c}{SD} & \multicolumn{1}{c}{N} & \multicolumn{1}{c}{Mean} & \multicolumn{1}{c|}{SD} \\ 
\hline
Mental component summary & 1043 & 43.78 & 11.8 & 466 & 42.74 & 12.4 & 881 & 45.44 & 10.86 & 264 & 44.44 & 11.4 \\ 
Employed full-time & 1043 & 0.31 & 0.46 & 466 & 0.33 & 0.47 & 881 & 0.39 & 0.49 & 264 & 0.51 & 0.5 \\ 
Employed part-time & 1043 & 0.25 & 0.43 & 466 & 0.25 & 0.43 & 881 & 0.31 & 0.46 & 264 & 0.23 & 0.42 \\ 
Neighbourhood disadvantage quintile & 1043 & 2.5 & 1.36 & 466 & 2.47 & 1.27 & 881 & 2.76 & 1.34 & 264 & 2.78 & 1.28 \\ 
Disposable income (\$1,000's) & 1043 & 37.56 & 17.88 & 466 & 36.98 & 15.28 & 881 & 45.15 & 22.97 & 264 & 47.16 & 21.93 \\ 
Physical component summary & 1043 & 50.05 & 10.26 & 466 & 48.94 & 10.88 & 881 & 50.55 & 10.13 & 264 & 49.38 & 9.93\\ 
\hline
\hline
\end{tabular}
}
\end{table}


	\section{Empirical Results}\label{sec.results}

		\subsection{Baseline Results}


The FE models in equation (\ref{eq.fe.estimation1}) and equation (\ref{eq.fe.estimation2}) are estimated for both renters in the private market, and for owners with a mortgage, using the full HILDA dataset. After removing missing observations, estimation is performed with 12,644 individuals for the owners' regressions, and 10,046 individuals for the renters' regressions. The mental health dependent variable is represented using the MCS, and the coefficient representing the additional effect from housing stress, the key variable of interest, is estimated from equation (\ref{eq.fe.estimation2}). In order to explore the impact of omitting financial hardship from the mental health equation, estimations of equation (\ref{eq.fe.estimation1}) are provided for comparison. To assess the individual contribution of each financial hardship indicator, FE regressions of the MCS on each of the eight financial hardship indicators are also included. Control variables are used in all the regressions, along with time FEs (using calendar year dummy variables).

The baseline results are presented in Table \ref{tab.mcs.fe.baseline}, with the renters' regressions in columns (1) to (3), and the owners' regressions in columns (4) to (6). The estimated coefficients for equation (\ref{eq.fe.estimation1}) are presented in columns (1) and (4), with the estimates for equation (\ref{eq.fe.estimation2}) in columns (2) and (5). The results for including each financial hardship variable separately are found in columns (3) and (6). Cluster robust standard errors are reported in parentheses, with clustering at the individual level. The R-squared reported relates to the within variation only (i.e. the variation remaining after the demeaning has been performed). The F-statistic reported is for the null hypothesis that the joint effect of all financial hardship variables together is zero.

\begin{table}[ht]
\centering
\caption{Baseline Regression Results}

\begingroup 
\footnotesize 
\begin{tabular}{@{\extracolsep{2pt}}lcccccc} 
\multicolumn{7}{l}{\textit{Dependent variable: Mental Component Summary}} \\
\\[-1.8ex]\hline 
\hline \\[-1.8ex] 
 & \multicolumn{3}{c}{Private Renters} & \multicolumn{3}{c}{Owners with Mortgage} \\ 
\\[-1.8ex] & (1) & (2) & (3) & (4) & (5) & (6)\\ 
\hline \\[-1.8ex] 
 Housing stress & $-$2.27$^{***}$ & $-$1.79$^{***}$ & $-$1.01$^{***}$ & $-$1.52$^{***}$ & $-$0.86$^{***}$ & $-$0.50$^{*}$ \\ 
  & (0.19) & (0.20) & (0.20) & (0.20) & (0.21) & (0.20) \\ 
  Financial hardship &  & $-$1.43$^{***}$ &  &  & $-$1.18$^{***}$ &  \\ 
  &  & (0.13) &  &  & (0.11) &  \\ 
  Could not pay bills &  &  & $-$0.98$^{***}$ &  &  & $-$0.61$^{***}$ \\ 
  &  &  & (0.17) &  &  & (0.15) \\ 
  Sold/pawned something &  &  & $-$0.85$^{***}$ &  &  & $-$1.01$^{***}$ \\ 
  &  &  & (0.22) &  &  & (0.24) \\ 
  Went without meals &  &  & $-$1.92$^{***}$ &  &  & $-$1.75$^{***}$ \\ 
  &  &  & (0.28) &  &  & (0.39) \\ 
  Went without heating &  &  & $-$1.46$^{***}$ &  &  & $-$1.02$^{**}$ \\ 
  &  &  & (0.31) &  &  & (0.37) \\ 
  Sort help from friends/family &  &  & $-$0.64$^{***}$ &  &  & $-$0.82$^{***}$ \\ 
  &  &  & (0.15) &  &  & (0.15) \\ 
  Sort help from community &  &  & $-$0.53$^{*}$ &  &  & $-$1.15$^{***}$ \\ 
  &  &  & (0.25) &  &  & (0.35) \\ 
  Cannot raise emergency money &  &  & $-$0.97$^{***}$ &  &  & $-$0.91$^{***}$ \\ 
  &  &  & (0.17) &  &  & (0.19) \\ 
 \hline \\[-1.8ex] 
Controls & Yes & Yes & Yes & Yes & Yes & Yes \\ 
Individual FEs & Yes & Yes & Yes & Yes & Yes & Yes \\
Year dummies & Yes & Yes & Yes & Yes & Yes & Yes \\ 
Individuals & 10046 & 10046 & 10046 & 12644 & 12644 & 12644 \\ 
Degrees of freedom & 31095 & 31094 & 31088 & 57135 & 57134 & 57128 \\ 
R-Squared & 0.0118 & 0.0161 & 0.0261 & 0.0045 & 0.0072 & 0.0108 \\ 
F-Statistic & 142.46 & 134.33 & 53.46 & 60.18 & 83.38 & 27.712 \\ 
\hline 
\hline \\[-1.8ex] 
\multicolumn{7}{r}{$^{*}$p$<$0.05; $^{**}$p$<$0.01; $^{***}$p$<$0.001} \\ 
\end{tabular} 
\endgroup 

\floatfoot{\textit{Notes:} Individual level cluster robust standard errors are reported in the parentheses. The F-statistic null hypothesis is all financial hardship variables are jointly zero. R-Squared values are calculated using within variation only. Control variables used are the natural logarithm of equivalised household disposable income, employment status (full-time, part-time, other), a categorical variable for the neighbourhood quintile of disadvantage of residence, and the first lag of the PCS score. Disposable income (in constant 2019 dollars) is at the household level and is equivalised based on the number of members following OECD equivalence weights.}
\label{tab.mcs.fe.baseline}
\end{table}
Before focusing on the estimated additional effect from housing stress, it is first worth noting that most coefficients have the expected sign (refer to Table \ref{tab.mcs.fe.baseline.full} in appendix \ref{sec.full.reg.results} for coefficient estimates of the full model, including the control variables). The effect from full or part-time employment on mental health is positive compared to the baseline case of no employment; increases in disposable income and improvements in physical health both have positive impacts on mental health; while living in a less disadvantaged neighbourhood tends to have positive impacts, but not necessarily so.


The \textit{Housing Stress} coefficient is the one of primary interest from Table \ref{tab.mcs.fe.baseline}. Let us first consider the FE estimates for renters. In column (1), when financial hardship is omitted from the equation, the \textit{Housing stress} coefficient is estimated at $-2.27$ and is highly significant.\footnote{The term highly significant is used for the $0.1$ percent level of significance.} Column (2) includes the financial hardship variable in the regression, reducing the magnitude of the \textit{Housing stress} coefficient, however it is still highly significant. This coefficient is estimated at $-1.79$, and has the following interpretation: Housing stress leads to a $0.18$ standard deviation decline in mental health, compared to non-housing related financial hardship alone. Thus, this is the additional affect of housing stress for an individual already in (non-housing related) financial hardship. The conversion to standard deviations uses the Australian population standard deviation for the MCS of 10. The \textit{Financial hardship} coefficient estimate of $-1.43$ is also highly significant, and similar, but slightly less, in magnitude to the \textit{Housing stress} estimate. This indicates that the coefficient in column (1) overestimates the effect of housing stress.

Renters in the private rental market that experience housing stress (and therefore also financial hardship) are estimated to have a total decline in the MCS of $(1.79 + 1.43 =)\, 3.22$ points on average, or $0.32$ standard deviations. In contrast, the effect on mental health of financial hardship without housing stress is estimated to be less than half this magnitude. 

Despite finding a strong additional effect of housing stress on renters, the estimated magnitude for owners is much smaller. The FE coefficient on \textit{Housing Stress} for owners, shown in column (5), is estimated to be $-0.86$, around half the magnitude found for renters, although it is still highly significant. As with renters, the magnitude of the \textit{Housing Stress} coefficient is dependent on inclusion or otherwise of \textit{Financial hardship}; with \textit{Financial hardship} omitted from the regression, the \textit{Housing Stress} coefficient estimate of $-1.52$ is almost double the magnitude. Interestingly, the impact of financial hardship is generally smaller in magnitude for owners, with the effect of non-housing related financial hardship alone estimated to be $-1.18$. The total effect of housing stress and financial hardship, an estimated $(0.86 + 1.18 =)\, 2.04$ point MCS decline, is less than two-thirds of that found for renters. 


Columns (3) and (5) of Table \ref{tab.mcs.fe.baseline} provide additional information as to the nature of financial hardship by including each of the eight hardship variables separately in a FE regression. These parameter estimates give an idea as to the possible contribution of each individual indicator towards the total effect of financial hardship on mental health. Interestingly, for renters (column (3)), while \textit{Housing stress} is significant, it is not the largest financial hardship coefficient; the deprivation variables, \textit{Went without meals} and \textit{Went without heating}, are especially large in magnitude. For example, the estimated coefficient on \textit{Went without meals} is $-1.92$, compared to $-1.01$ for \textit{Housing stress}. For owners (column (6)), the \textit{Housing stress} coefficient is the smallest of the eight hardship variables, estimated at $-0.50$, while \textit{Went without meals} is once again the largest, with an estimate of $-1.75$.

	\subsection{Difference-in-differences for Matched Subsamples}

\begin{figure}[!ht]
\centering
\caption{Matched Difference-in-differences Results}
\includegraphics[scale=.6]{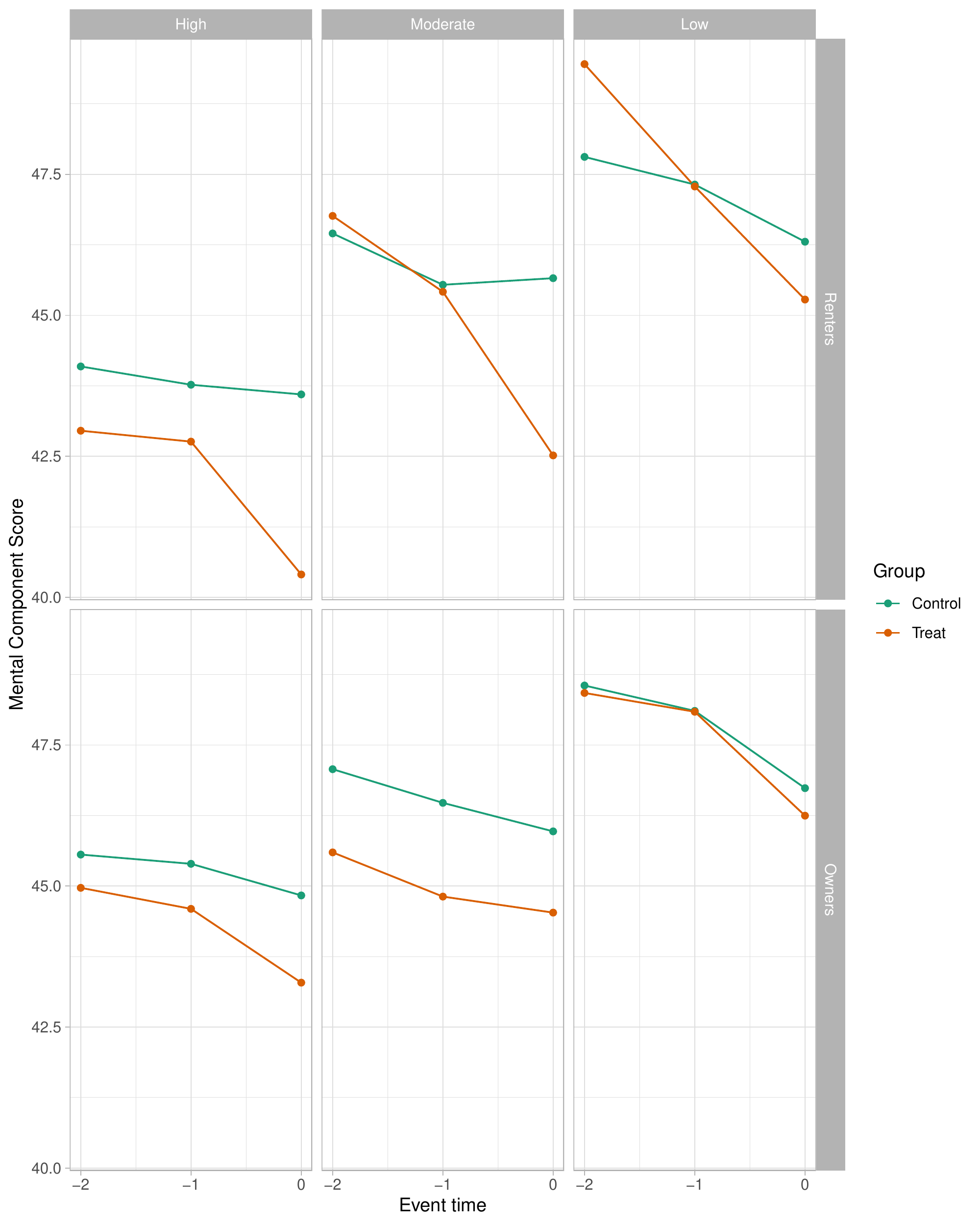}
\label{fig.results.match.trends.plot.rent}
\floatfoot{\textit{Notes:} The predicted MCS values plotted are conditional on event time and treatment group, and at the mean of control variables for the given time period. The treatment group has housing related financial hardship and non-housing related financial hardship at $t=0$, whereas the control group has non-housing related financial hardship only at $t=0$. All individuals in the \textit{High} subsamples (left column) experience non-housing related financial hardship prior in the pre-treatment periods; all individuals in the \textit{Low} subsamples (right column) are free of financial hardship in the pre-treatment periods; and individuals the \textit{Moderate} subsamples (middle column) have non-housing related financial hardship in one pre-treatment period only (at $t=-1$).}
\end{figure}

The DID methodology discussed in Section \ref{sec.het.causal.effects} attempts to improve the robustness to omitted variables and reverse causality by incorporating matching based on multiple periods, and allowing for explicit testing of the assumptions required for identification. Additionally, it permits us to investigate the possibility of heterogeneous effects. These results are summarised in the two-by-three grid presented in Figure \ref{fig.results.match.trends.plot.rent}. The rows differentiate based on tenure, with renters on the top row and owners on the bottom row. DID estimates for each tenure are produced for \textit{High}, \textit{Moderate}, and \textit{Low} subsamples. All individuals in the \textit{High} subsamples (left column) experience non-housing related financial hardship prior in the two pre-treatment periods; all individuals in the \textit{Low} subsamples (right column) are free of financial hardship in the two pre-treatment periods; and individuals the \textit{Moderate} subsamples have non-housing related financial hardship in one pre-treatment period only (at $t=-1$). It is hypothesised that non-housing related financial hardship in the pre-treatment period will increase the negative effect of housing related financial hardship on mental health, compared to a pre-treatment period free from financial hardship. 

The predicted MCS scores in figure \ref{fig.results.match.trends.plot.rent} are based on OLS estimation of the DID model in equation (\ref{eq.did.estimation}) for each subsample. The predictions shown are conditional on a given treatment or control group status and event time period; the control variables are predicted at the mean for a given time period, for the all observations in the subsample. Thus, the plots represent the treatment and control group differences in the estimates of the key coefficients of interest (see table \ref{tab.did.full} in appendix \ref{sec.full.reg.results} for the full set of coefficient estimates). At event time $t=0$, the treatment group (in red) experiences housing stress in addition to non-housing related financial hardship, whereas the control group (green) experiences non-housing related financial hardship only. Differences in the treatment and control group time trends from $t=-1$ to $t=0$ are attributed to the effect of the treatment, given identification assumptions hold.


Prior to discussing the causal estimates of housing stress, we first investigate the validity of the key parallel trends assumptions by examining the pre-treatment trends from $t=-2$ to $t=-1$. Inspection of Figure \ref{fig.results.match.trends.plot.rent} show the trends in the pre-treatment periods are indeed close to parallel for treatment and control groups in five of the six panels; the exception being the \textit{Low} subgroup for renters. The parallel trends assumption can be formally tested with a t-test of $\gamma_2=0$ from equation (\ref{eq.did.estimation}), representing the null hypothesis that pre-treatment trends are parallel. The results for each subgroup are in the top row of Table \ref{tab.het.results.rent.own}, labelled \textit{Parallel Trends}. The null of parallel trends is rejected at the 10 percent level of significance for the \textit{Low} subsample for renters, with a p-value of $0.08$. However, for the other five subsamples, the corresponding p-values are large and there is no evidence to reject the parallel trends assumption.

\begin{table}
\centering
\caption{Difference-in-differences Hypothesis Testing}
\begingroup\normalsize
\begin{tabular}{llrrrrrr}
  \hline
   & & \multicolumn{3}{c}{Renters} & \multicolumn{3}{c}{Owners} \\
Test &  & \hspace{3mm} Low &  Mod &  High  & \hspace{3mm} Low & Mod & High \\ 
  \hline
\hline
Parallel Trends & Estimate & -1.68 & -0.43 & 0.13 & 0.12 & -0.19 & -0.21 \\ 
   & p-value & 0.08 & 0.70 & 0.88 & 0.87 & 0.88 & 0.83 \\ 
   \hline
Treatment Effects & Estimate & -2.67 & -3.45 & -2.05 & -0.36 & 0.03 & -0.96 \\ 
   & p-value & 0.02 & 0.00 & 0.02 & 0.66 & 0.98 & 0.31 \\ 
   \hline
\hline
\end{tabular}
\endgroup

\floatfoot{\textit{Notes:} All tests are performed with t-tests of equation (\ref{eq.did.estimation}) using individual level cluster robust standard errors. Parallel trends is tested with a null of $\gamma_{2}^{(h)} = 0$. Treatment effects are tested with a null of $\tau^{(h)} = 0$.}
\label{tab.het.results.rent.own}
\end{table}

Having shown the appropriateness of the key identification assumption in five of the six cases, we can focus on the estimated causal effects from the matched DID, ignoring the \textit{Low} subgroup results for renters. Visually inspecting the \textit{High} subgroup for renters on the top left of Figure \ref{fig.results.match.trends.plot.rent}, the parallel trends that hold in the pre-treatment periods are seen to diverge from $t=-1$ to $t=0$, with a greater downward trend in the treatment group. The estimated DID treatment effects are shown in Table \ref{tab.het.results.rent.own}. For renters in the \textit{High} subgroup, the estimated additional effect of housing related financial hardship is $-2.05$, which is significant at the 5 percent significance level.
 
The \textit{Moderate} subgroup for renters, seen in the top middle panel of figure \ref{fig.results.match.trends.plot.rent}, also shows the parallel trends deviating from $t=-1$ to $t=0$. However, some caution is required in interpreting these results as the power to reject the parallel trends is lower than in the \textit{High} subgroup. The estimate of $\gamma_2$ for the \textit{Moderate} subgroup is $-0.43$, with a standard deviation of $1.13$, compared to an estimate of $0.13$ and standard deviation of $0.84$ for the \textit{High} subgroup. The larger standard deviation is due to the small subsample size; as can see in Table \ref{tab.ss.subgroups.mod} and Table \ref{tab.ss.subgroups.high}, there are only $125$ observations in the \textit{Moderate} treatment group, compared to $466$ observations for the \textit{High} treatment group.\footnote{This sample size problem also illustrates why only two pre-treatment periods are used, as discussed briefly in Section \ref{sec.matchss}.} Therefore, we should be cautious about placing too much weight on the effect of housing stress in the \textit{Moderate} subgroup, estimated to be $-3.45$. However, this result is still qualitatively consistent with a finding that housing stress has significant effects on renters, especially for those in prior financial hardship.

The bottom row of figure \ref{fig.results.match.trends.plot.rent} shows the estimated DID results for owners. Visually, each panel reveals the trends from $t=-1$ to $t=0$ are similar for the treatment and control groups. Table \ref{tab.did.full} shows the treatment effects, which are estimated to be $-0.36$, $0.03$, and $-0.96$ for the \textit{Low}, \textit{Moderate}, and \textit{High} subsamples; these estimates are not significantly different from zero. 

The results from the DID estimations reveal a consistency with the FE results. Both methodologies show the impact of housing stress depends on tenure. For renters, the FE results indicate housing stress has a strong additional effect on mental health, compared to non-housing related financial hardship alone. The DID results indicate the same conclusion for individuals who are already in high financial stress, indicated by non-housing related financial hardship in the two pre-treatment periods. The estimated additional effect of $-2.05$ is slightly higher than the FE estimate of $-1.79$. For owners, the small impact reported in the FE regressions is not found in the DID results. This suggests there is little to no additional impact on mental health of housing stress for owners.


	\subsection{Robustness}
	
	\subsubsection{Assessing Common Support}
		
The DID results were estimated with matched subgroups based on the matching rules outlined in Section \ref{sec.het.causal.effects}. The parallel trends observed in five of the six panels of Figure \ref{fig.results.match.trends.plot.rent} provide confidence that the matched treatment and control groups adequately represent counterfactual outcomes. This is explored further by considering the similarity between the treatment and control groups based on observed covariate values. Besides parallel trends, formal proofs of identification in the DID literature also rely on the so-called common support assumption \citep{lechner2011estimation}. Matching methodologies, such as propensity score matching, typically assess the common support assumption by comparing distributions of propensity scores between treatment and control groups; the same assessment is performed on the matched subsamples used in this research.

Figure \ref{fig.results.match.psm.plot.rent} in the appendix shows the assessment of the common support assumption, with the rows showing the results for each of the six subsamples. For a given subsample and event time, the propensity score is estimated for each observation with a standard logistic regression using the covariates from the DID analysis.\footnote{The model used to estimate the propensity scores is $p(W)=logistic(W^{\prime} \theta)$ where $W$ consists of the natural logarithm of equivalised household disposable income, employment status (full-time, part-time, other), an indicator for the neighbourhood quintile of disadvantage, an indicator for the calendar year the event occurs, and the lag of PCS.} The distribution of the predicted propensity scores for the control and treatment groups is then approximated using histograms, and plotted for comparability. Each event time period is assessed separately, with the columns showing event time periods $t=-2, t=-1,$ and $t=0$.


The histograms from Figure \ref{fig.results.match.psm.plot.rent}, representing the probability of being in the treatment group for each observation, show a high degree of symmetry between the treatment and control groups. This implies adequate overlap in the covariate distribution for the treatment and control groups, satisfying the matching requirement.


\begin{figure}[ht]
\centering
\caption{Assessment of Common Support Assumption}
\includegraphics[scale=.69]{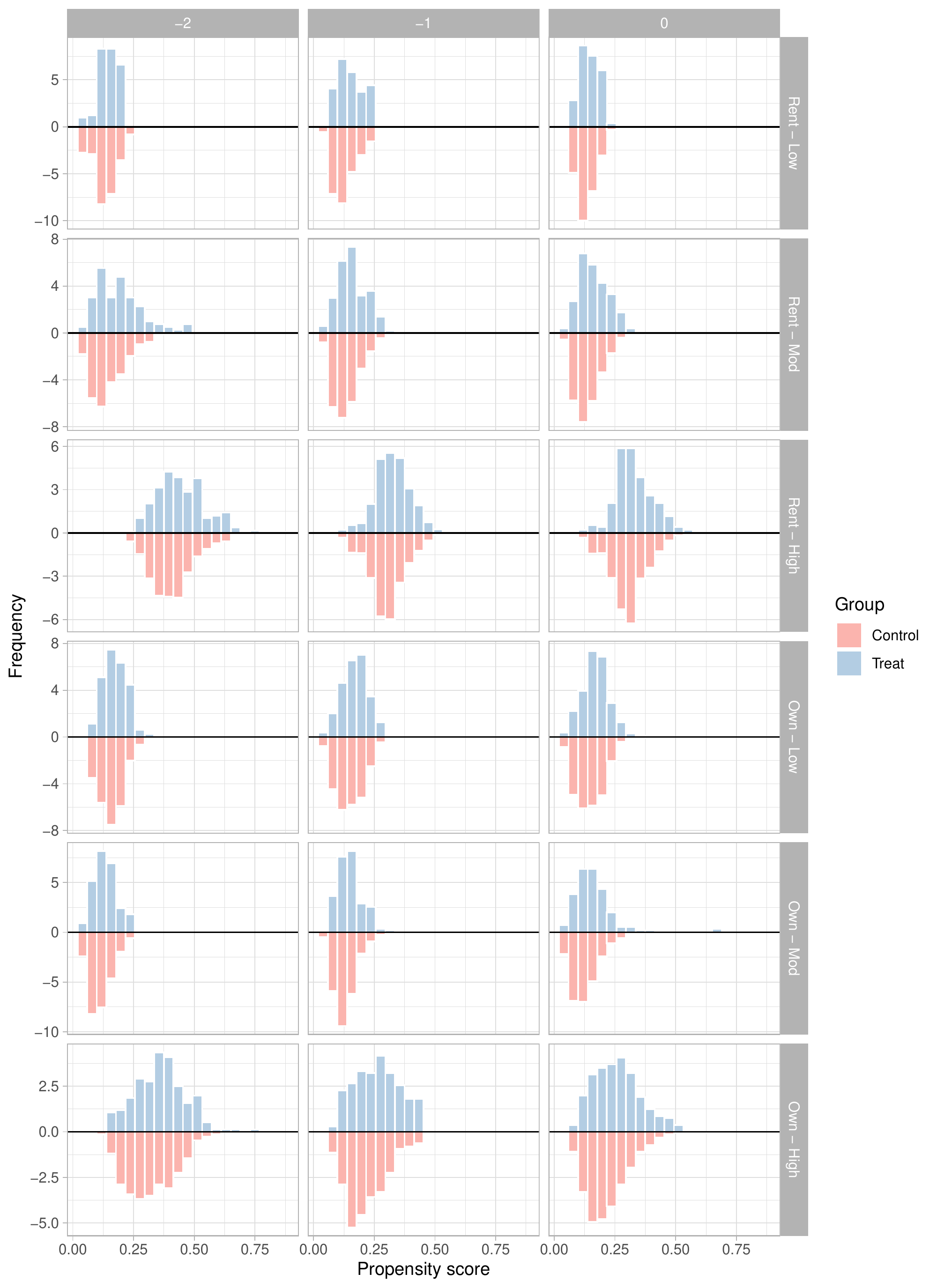}
\label{fig.results.match.psm.plot.rent}
\floatfoot{\textit{Notes:} The propensity score is the probability of being in the treatment group for a given event time period, estimated using logistic regression using covariates: the natural logarithm of equivalised household disposable income, employment status (full-time, part-time, other), an indicator for the neighbourhood quintile of disadvantage, an indicator for the calendar year the event occurs, and the lag of PCS. Disposable income (in constant 2019 dollars) is at the household level and is equivalised based on the number of members following OECD equivalence weights.}
\end{figure}

	\subsubsection{Alternative Dependent Variable}
	
The sensitivity of the FE and DID results to the choice of dependent variable is also assessed. An alternative outcome measure is available from the SF-36 called the mental health scale. The mental health scale is a narrower measure of mental health compared to the MCS. It is constructed from summing the responses to five mental health questions asked in the SF-36, and transforming the scale to range from 0 to 100. The MCS represents a broader measure of mental health as it weights responses to additional questions regarding vitality, social functioning, and role limitations due to emotional functioning.

Using the mental health scale as the dependent variable, the results are re-estimated, with the robustness results for the FE regressions provided in Table \ref{tab.mcs.fe.baseline.sf36mh} in the appendix, and the DID results in figure \ref{fig.results.match.trends.plot.rent.sf36mh} and Table \ref{tab.het.results.did.sum.sf36mh}, also in the appendix. The FE results show a similar direction, magnitude, and significance as the main reported results. The coefficient magnitudes can be compared using the mental health scale population standard deviation for Australia of 16.96 \citepalias{australian1997national}. For example, using the mental health scale the total effect of housing stress for renters is a $0.28$ standard deviation decline, compared to a $0.32$ standard deviation decline found in the main results. 

Figure \ref{fig.results.match.trends.plot.rent.sf36mh} shows a consistency with the main DID results. Visually, the \textit{Low} subgroup for rents once again does not meet the parallel trends assumption, and caution is also required for the \textit{High} and \textit{Moderate} subgroups for owners. Despite these precautions, the qualitative results are very similar to the main DID findings; there is evidence of an additional effect on mental health from housing stress for renters with financial hardship in the pre-treatment periods, but little evidence of housing related effects for owners.

	\section{Discussion}\label{sec.discussion}

	\subsection{Interpretation of Results and Limitations}	
	
The FE results have estimated the additional effect of housing stress on mental health, compared to non-housing related financial hardship alone, after controlling for income, employment, physical health, neighbourhood effects, and individual and time FEs. The estimation of equation \ref{eq.fe.estimation2} found a decline of $0.18$ standard deviations in the MCS. We saw in the FE regressions that financial hardship also had an impact on mental health, and omitting this variable biases the magnitude of the \textit{Housing stress} coefficient upwards. This highlights the importance of including financial hardship when estimating the effect of housing stress. The DID framework explicitly matched on financial hardship experience over multiple periods and allowed for testing of the key identification assumption (that of parallel trends). There is reassurance that two different methodologies produce similar conclusions. 

	
The results from the FE regressions suggest the effect on mental health from housing related financial hardship depends on tenure, with renters more severely impacted than owners. The DID results suggest that the individuals most affected by housing stress are renters already exposed to financial hardship, and owners could be unaffected. These differences based on tenure could be explained by income and wealth differences. In the sample used for analysis, owners have higher median incomes, with equivalised disposable household incomes almost \$12 thousand more than renters in 2018.\footnote{The median of equivalised disposable income across the entire sample in 2018 is \$49 thousand.} The median net wealth for a renter is \$66 thousand in 2018, compared to \$919 thousand for owners. Owners with missed mortgage repayments still have a median net wealth of \$432 thousand, multiple times more than the median for renters.\footnote{It is worth noting that owners are on average four years older than renters, however this can only explain some of these differences.} Considering income and wealth disparities by tenure, there are likely to be more serious implications for renters in housing stress, compared with owners.


	Although the results suggest renters face additional mental health impacts from housing stress, they do not necessarily mean that housing stress has a larger impact than other aspects of financial hardship, despite the specific features of housing expenditure discussed earlier.
	The FE regressions found that non-housing related financial hardship also has significant negative impacts on mental health. Additionally, a regression that simply adds each of the eight categories of financial hardship as separate variables suggests all aspects of financial hardship impact on mental health, and that deprivation, such as going without meals or heating, could have a larger impact than housing stress. Further conclusions are limited by the simultaneous nature by which financial hardship indicators are determined. For example, although deprivation appears to have large mental health impacts, burdensome housing costs could be a cause of this deprivation. 



A limitation of this study is that, despite highly significant results, the FE model only describes a small amount of the within variation used for estimation. Leaving a large fraction of the variance unexplained increases the possibility of omitted variable bias. Reverse causality is also a concern in the FE regressions. For example, increasingly erratic or impulsive behaviour associated with worsening mental health could cause financial hardship problems, or missed housing payments. This problem is mitigated to some degree by the timing of these variables. The HILDA Survey questions for the MCS are asked in relation to the previous month, whereas seven of the eight financial hardship indicators refer to events that have occurred since the start of the interview year.\footnote{The majority of the HILDA interviews take place between July and October of each year.} The FE results also use the first lag of the PCS to reduce the problem of simultaneously determined mental and physical health.

Although the DID has the advantage over the FE regressions of matching over multiple periods, and being able to test the identification assumption, once matching is performed there are a limited number of observations in some subsamples. This reduces the ability to reject the parallel trends assumption, and increases the standard errors around the parameter estimating the effect size. Despite these concerns, the similarity of the FE and DID results provide reassurance as to the robustness of the main findings. 

	\subsection{The Broader Literature and Policy Implications}
		

This research fits within a broader field that seeks a better understanding of the social determinates of health, which include income, education, housing, and others \citep{mikkonen2010social}. This area has gained in importance due to the inability for traditional explanations, such has individual health behaviours, poverty, and access to health care, to fully explain mortality, morbidity, and other health gradients, including mental health, that are found in most countries across socioeconomic status \citep{marmot2014social, smith1999healthy}. There are a number of reasons to suspect that social constructs that contribute to socioeconomic status are also important causes of the health gradient. These reasons include chronic arousal of stress pathways leading to physiological disregulation, and increases in psychosocial stressors due to positioning within the social hierarchy \citep{deaton2001mortality, mcewen2010central}. 

One criticism of the literature on the social determinant of health is a lack of emphasis on strong causal methodology. And due to the highly correlated nature of many of these variables, such as income, education and wealth, the exact causal pathways are difficult to ascertain. For example, unequal health outcomes in adults attributed to income could be caused by health shocks early in life that reduce lifetime earnings \citep{smith1999healthy}. In order to correctly inform policy makers, more studies with a causal focus are needed. This research contributes to the literature by addressing methodological concerns in the literature investigating the effect of housing stress on mental health.

The results presented in this research provide support for government intervention to alleviate housing stress for renters, especially those in persistent financial hardship, in order to reduce the negative impacts on mental health. Increasing the supply of affordable housing options, or increasing rental supplements, would both appear to have mental health benefits. However, it is worth noting that the evidence presented in this research does not necessarily favour housing interventions over other welfare policies that may reduce financial hardship more broadly. For instance, the results suggest that experiences of deprivation could have especially large negative mental health impacts.



The effectiveness of policy to reduce the impact of housing stress and financial hardship on mental health may benefit from looking beyond the usual policy prescriptions. Inspired by behavioural economics, some researchers have started asking how economic decisions change under financial hardship. For example, \cite{carvalho2016poverty} finds that individuals are more present bias before payday than after payday. Potentially short term financial problems can become overly salient in the minds of decision makers, leading to behaviours that result in a long term poverty trap. This appears to open up new policy directions for improving financial mobility, and therefore mental health, such as interventions to help initial asset accumulation that could assist with escaping a poverty trap \citep{bernheim2015poverty}.

	\section{Conclusion}\label{sec.conclusion}

Housing expenditure tends to be sticky and costly to adjust, and makes up a large proportion of household expenditure. Additionally, the loss of housing can have catastrophic consequences. These specific features of housing expenditure imply that housing stress could cause negative mental health impacts. This research finds that housing stress does have an effect on mental health, with larger declines estimated for renters than for owners. Especially impacted are renters that have prior experience with financial hardship. For these individuals, the DID estimates a $0.2$ standard deviation decline in the MCS. This is the additional impact of housing stress, compared to non-housing related financial hardship alone. The estimated impact of housing stress on owners with a mortgage is small or not different from zero. This research also suggests that the mental health impact of housing stress is more important than some, but not all, aspects of financial hardship.  


\newpage
\bibliographystyle{newapa}
\bibliography{ch2_draft2}
\newpage

	\section{Appendix}

	\subsection{Full Regression Results} \label{sec.full.reg.results}

\begin{table}[!h]
\centering
\vspace{-4mm}
\caption{Baseline Regression Results - Full Model}

\begingroup 
\footnotesize 
\begin{tabular}{@{\extracolsep{2pt}}lcccccc} 
\\[-1.8ex]\hline 
\hline \\[-1.8ex] 
 & \multicolumn{6}{c}{\textit{Dependent variable:}} \\ 
\cline{2-7} 
\\[-1.8ex] & \multicolumn{6}{c}{Mental Component Summary} \\ 
 & \multicolumn{3}{c}{Renters} & \multicolumn{3}{c}{Owners} \\ 
\\[-1.8ex] & (1) & (2) & (3) & (4) & (5) & (6)\\ 
\hline \\[-1.8ex] 
 Housing stress & $-$2.27$^{***}$ & $-$1.79$^{***}$ & $-$1.01$^{***}$ & $-$1.52$^{***}$ & $-$0.86$^{***}$ & $-$0.50$^{*}$ \\ 
  & (0.19) & (0.20) & (0.20) & (0.20) & (0.21) & (0.20) \\ 
  Financial hardship &  & $-$1.43$^{***}$ &  &  & $-$1.18$^{***}$ &  \\ 
  &  & (0.13) &  &  & (0.11) &  \\ 
  Could not pay bills &  &  & $-$0.98$^{***}$ &  &  & $-$0.61$^{***}$ \\ 
  &  &  & (0.17) &  &  & (0.15) \\ 
  Sold/pawned something &  &  & $-$0.85$^{***}$ &  &  & $-$1.01$^{***}$ \\ 
  &  &  & (0.22) &  &  & (0.24) \\ 
  Went without meals &  &  & $-$1.92$^{***}$ &  &  & $-$1.75$^{***}$ \\ 
  &  &  & (0.28) &  &  & (0.39) \\ 
  Went without heating &  &  & $-$1.46$^{***}$ &  &  & $-$1.02$^{**}$ \\ 
  &  &  & (0.31) &  &  & (0.37) \\ 
  Sort help from friends/family &  &  & $-$0.64$^{***}$ &  &  & $-$0.82$^{***}$ \\ 
  &  &  & (0.15) &  &  & (0.15) \\ 
  Sort help from community &  &  & $-$0.53$^{*}$ &  &  & $-$1.15$^{***}$ \\ 
  &  &  & (0.25) &  &  & (0.35) \\ 
  Cannot raise emergency money &  &  & $-$0.97$^{***}$ &  &  & $-$0.91$^{***}$ \\ 
  &  &  & (0.17) &  &  & (0.19) \\ 
  Household Disposable income & 0.22 & 0.14 & 0.07 & 0.53$^{***}$ & 0.46$^{***}$ & 0.42$^{***}$ \\ 
  & (0.12) & (0.12) & (0.12) & (0.11) & (0.11) & (0.11) \\ 
  Employed full-time & 1.54$^{***}$ & 1.36$^{***}$ & 1.18$^{***}$ & 0.85$^{***}$ & 0.78$^{***}$ & 0.73$^{***}$ \\ 
  & (0.20) & (0.20) & (0.20) & (0.16) & (0.16) & (0.16) \\ 
  Employed part-time & 1.24$^{***}$ & 1.17$^{***}$ & 1.07$^{***}$ & 0.83$^{***}$ & 0.80$^{***}$ & 0.77$^{***}$ \\ 
  & (0.19) & (0.19) & (0.19) & (0.15) & (0.15) & (0.15) \\ 
  Physical component summary & 0.04$^{***}$ & 0.04$^{***}$ & 0.04$^{***}$ & 0.02$^{***}$ & 0.02$^{***}$ & 0.02$^{***}$ \\ 
  & (0.01) & (0.01) & (0.01) & (0.01) & (0.01) & (0.01) \\ 
  Neighbourhood (Q2) & 0.11 & 0.13 & 0.10 & 0.29 & 0.31 & 0.29 \\ 
  & (0.23) & (0.23) & (0.22) & (0.30) & (0.30) & (0.30) \\ 
  Neighbourhood (Q3) & $-$0.12 & $-$0.13 & $-$0.15 & 0.09 & 0.11 & 0.11 \\ 
  & (0.24) & (0.24) & (0.23) & (0.29) & (0.29) & (0.29) \\ 
  Neighbourhood (Q4) & 0.12 & 0.13 & 0.08 & $-$0.32 & $-$0.29 & $-$0.31 \\ 
  & (0.25) & (0.25) & (0.25) & (0.30) & (0.30) & (0.30) \\ 
  Neighbourhood (Q5) & 0.22 & 0.22 & 0.18 & $-$0.02 & 0.004 & $-$0.01 \\ 
  & (0.28) & (0.28) & (0.28) & (0.33) & (0.33) & (0.33) \\ 
 \hline \\[-1.8ex] 
Year dummies & Yes & Yes & Yes & Yes & Yes & Yes \\ 
Individuals & 10046 & 10046 & 10046 & 12644 & 12644 & 12644 \\ 
Degrees of freedom & 31095 & 31094 & 31088 & 57135 & 57134 & 57128 \\ 
\hline 
\hline \\[-1.8ex] 
\multicolumn{7}{r}{$^{*}$p$<$0.05; $^{**}$p$<$0.01; $^{***}$p$<$0.001} \\ 
\end{tabular} 
\endgroup 

\floatfoot{\textit{Notes:} Cluster robust standard errors (individual level) reported in the parentheses. R-Squared values are calculated using within variation only. Income is natural logarithm of equivalised household disposable income (constant 2019 dollars), equivalised by household members using OECD equivalence weights. The PCS is lagged by one period.}
\label{tab.mcs.fe.baseline.full}
\end{table}

\begin{table}[!h]
\centering
\caption{Difference-in-differences Results - Full Model}

\begingroup 
\small 
\begin{tabular}{@{\extracolsep{5pt}}lcccccc} 
\\[-1.8ex]\hline 
\hline \\[-1.8ex] 
 & \multicolumn{6}{c}{\textit{Dependent variable:}} \\ 
\cline{2-7} 
\\[-1.8ex] & \multicolumn{6}{c}{Mental Component Summary} \\ 
 & \multicolumn{3}{c}{Renters} & \multicolumn{3}{c}{Owners} \\ 
\\[-1.8ex] & (1) & (2) & (3) & (4) & (5) & (6)\\ 
\\[-1.8ex] & Low & Mod & High & Low. & Mod. & High.\\ 
\hline \\[-1.8ex] 
 Treat:1[t=0] & $-$2.67$^{*}$ & $-$3.45$^{**}$ & $-$2.05$^{*}$ & $-$0.36 & 0.03 & $-$0.96 \\ 
  & (1.11) & (1.21) & (0.87) & (0.81) & (1.35) & (0.94) \\ 
  Treat:1[t=-1] & $-$1.68 & $-$0.43 & 0.13 & 0.12 & $-$0.19 & $-$0.21 \\ 
  & (0.97) & (1.13) & (0.84) & (0.71) & (1.21) & (0.95) \\ 
  Treat & 1.64 & 0.31 & $-$1.14 & $-$0.13 & $-$1.47 & $-$0.59 \\ 
  & (0.99) & (1.09) & (0.86) & (0.75) & (1.27) & (0.96) \\ 
  1[t=-1] & $-$0.61 & $-$0.93$^{*}$ & $-$0.78 & $-$0.47 & $-$0.82$^{*}$ & $-$0.15 \\ 
  & (0.37) & (0.47) & (0.53) & (0.29) & (0.42) & (0.52) \\ 
  1[t=0] & $-$1.56$^{***}$ & $-$0.87 & $-$0.96 & $-$1.82$^{***}$ & $-$1.39$^{**}$ & $-$0.74 \\ 
  & (0.41) & (0.45) & (0.55) & (0.32) & (0.43) & (0.54) \\ 
  Disposable income & $-$0.31 & $-$1.47$^{*}$ & $-$0.10 & 0.43 & $-$0.42 & $-$0.06 \\ 
  & (0.54) & (0.60) & (0.48) & (0.41) & (0.64) & (0.57) \\ 
  Employed full-time & 1.76$^{*}$ & 3.00$^{***}$ & 2.77$^{***}$ & 2.01$^{***}$ & 3.44$^{***}$ & 2.95$^{***}$ \\ 
  & (0.69) & (0.72) & (0.61) & (0.59) & (0.78) & (0.73) \\ 
  Employed part-time & 0.55 & 0.05 & 2.18$^{***}$ & 1.73$^{**}$ & 2.69$^{***}$ & 2.15$^{**}$ \\ 
  & (0.72) & (0.78) & (0.62) & (0.61) & (0.80) & (0.74) \\ 
  Physical component summary & 0.002 & $-$0.02 & 0.06$^{*}$ & 0.06$^{*}$ & 0.02 & 0.07$^{*}$ \\ 
  & (0.03) & (0.03) & (0.03) & (0.02) & (0.03) & (0.03) \\ 
  Neighbourhood (Q2) & $-$0.74 & 0.28 & 1.66$^{*}$ & 0.02 & 0.90 & $-$0.76 \\ 
  & (0.77) & (0.87) & (0.65) & (0.65) & (0.86) & (0.81) \\ 
  Neighbourhood (Q3) & $-$0.65 & $-$0.30 & $-$0.07 & $-$0.74 & $-$0.12 & $-$1.10 \\ 
  & (0.81) & (0.92) & (0.77) & (0.64) & (0.92) & (0.87) \\ 
  Neighbourhood (Q4) & $-$0.25 & 1.87$^{*}$ & $-$0.31 & $-$0.98 & 1.00 & $-$1.17 \\ 
  & (0.86) & (0.91) & (0.86) & (0.66) & (0.94) & (0.92) \\ 
  Neighbourhood (Q5) & $-$1.70$^{*}$ & 1.04 & $-$0.03 & $-$1.50$^{*}$ & 0.56 & $-$0.45 \\ 
  & (0.86) & (0.96) & (0.92) & (0.67) & (0.94) & (0.92) \\ 
 \hline \\[-1.8ex] 
Event Year Fixed Effects & Yes & Yes & Yes & Yes & Yes & Yes \\ 
Individuals & 1078 & 928 & 1552 & 1774 & 1069 & 1183 \\ 
Observations & 2885 & 2441 & 3748 & 4698 & 2782 & 2869 \\ 
\hline 
\hline \\[-1.8ex] 
\multicolumn{7}{r}{$^{*}$p$<$0.05; $^{**}$p$<$0.01; $^{***}$p$<$0.001} \\ 
\end{tabular} 
\endgroup 

\floatfoot{\textit{Notes:} Cluster robust standard errors (individual level) reported in the parentheses. Income is natural logarithm of equivalised household disposable income (constant 2019 dollars), equivalised by household members using OECD equivalence weights. Colons represent interaction terms and $1[\cdot]$ are indicator variables for the stated event time. Event year fixed effects for the actual calendar year the event occurs, and are added using dummy variables for 2004-2019.\vspace{10mm}}

\label{tab.did.full}
\end{table}

		
	\newpage

		\subsection{MCS and PCS Construction} \label{sec.mcs.construct}
	
To construct the MCS and PCS, this research uses the ABS National Health Survey, 1995. This provides the SF-36 summary statistics for the Australian population, along with weighting coefficients derived from principal components factor analysis (see table \ref{tablepcaweights}). Following the procedure outlined in \cite{ware2001sf}, the eight category scores of the SF-36 are first standardised using the ABS's population mean and standard deviation estimates. The ABS weighting coefficients are then used to take linear combinations of the eight category scores to form mental and physical component summaries. In the final step, the two aggregate measures are transformed to have a mean of 50, and a standard deviation of 10 in the general population.
\\
\begin{table}[!htb]
\caption{MCS and PCS Creation}
\centering
\begin{tabular}{l c c c c}
\hline\hline
\text {SF-36 Scale} & \text { Pop. Mean } & \text { Pop. Std. Dev. } & \text { PCS Coeff. } & \text { MCS Coeff. } \\
\hline
\text { Physical function } & 83.46 & 23.23 & 0.47 & -0.24 \\
\text { Role limits (physical) } & 80.28 & 34.84 & 0.38 & -0.13 \\
\text { Bodily pain } & 76.94 & 24.84 & 0.37 & -0.12 \\
\text { General health } & 71.82 & 20.35 & 0.19 & 0.05 \\
\text { Vitality } & 64.48 & 19.77 & -0.02 & 0.27 \\
\text { Social function } & 85.06 & 22.29 & -0.01 & 0.26 \\
\text { Role limits (emotion) } & 83.19 & 32.15 & -0.15 & 0.36 \\
\text { Mental health } & 75.98 & 16.96 & -0.27 & 0.49 \\ 
\hline\hline
\end{tabular}\\
\floatfoot{Columns (left to right): Australian population means, Australian population standard deviations, weighting coefficients for PCS, weighting coefficients for MCS. Weighting coefficients are the principal components weighting coefficients for the eight scales of the SF-36 in the Australian population \citepalias{australian1997national}.}
\label{tablepcaweights}
\end{table}

\newpage
	\subsection{Robustness - Mental Health Scale Dependent Variable}\label{sec.robust.sf36mh}
	
\begin{table}[!hp]
\centering
\caption{Fixed Effects Robustness - Mental Health Scale Dependent Variable}

\begingroup 
\footnotesize 
\begin{tabular}{@{\extracolsep{2pt}}lcccccc} 
\multicolumn{7}{l}{\textit{Dependent variable: Mental Health Scale}}\\
\\[-1.8ex]\hline 
\hline \\[-1.8ex]  
 & \multicolumn{3}{c}{Private Renters} & \multicolumn{3}{c}{Owners with Mortgage} \\ 
\\[-1.8ex] & (1) & (2) & (3) & (4) & (5) & (6)\\ 
\hline \\[-1.8ex] 
 Housing stress & $-$3.32$^{***}$ & $-$2.54$^{***}$ & $-$1.46$^{***}$ & $-$2.52$^{***}$ & $-$1.44$^{***}$ & $-$0.85$^{**}$ \\ 
  & (0.30) & (0.31) & (0.31) & (0.31) & (0.33) & (0.32) \\ 
  Financial hardship &  & $-$2.29$^{***}$ &  &  & $-$1.93$^{***}$ &  \\ 
  &  & (0.21) &  &  & (0.18) &  \\ 
  Could not pay bills &  &  & $-$1.32$^{***}$ &  &  & $-$1.01$^{***}$ \\ 
  &  &  & (0.27) &  &  & (0.23) \\ 
  Sold/pawned something &  &  & $-$1.59$^{***}$ &  &  & $-$1.77$^{***}$ \\ 
  &  &  & (0.34) &  &  & (0.38) \\ 
  Went without meals &  &  & $-$2.87$^{***}$ &  &  & $-$2.19$^{***}$ \\ 
  &  &  & (0.44) &  &  & (0.61) \\ 
  Went without heating &  &  & $-$2.13$^{***}$ &  &  & $-$1.92$^{***}$ \\ 
  &  &  & (0.49) &  &  & (0.58) \\ 
  Sort help from friends/family &  &  & $-$0.86$^{***}$ &  &  & $-$1.46$^{***}$ \\ 
  &  &  & (0.23) &  &  & (0.22) \\ 
  Sort help from community &  &  & $-$0.80$^{*}$ &  &  & $-$1.79$^{**}$ \\ 
  &  &  & (0.40) &  &  & (0.55) \\ 
  Cannot raise emergency money &  &  & $-$1.57$^{***}$ &  &  & $-$1.48$^{***}$ \\ 
  &  &  & (0.28) &  &  & (0.30) \\ 
 \hline \\[-1.8ex] 
Controls & Yes & Yes & Yes & Yes & Yes & Yes \\ 
Year dummies & Yes & Yes & Yes & Yes & Yes & Yes \\ 
Individuals & 10046 & 10046 & 10046 & 12644 & 12644 & 12644 \\ 
Degrees of freedom & 31095 & 31094 & 31088 & 57135 & 57134 & 57128 \\ 
R-Squared & 0.0108 & 0.0152 & 0.0238 & 0.0046 & 0.0076 & 0.0112 \\ 
F-Statistic & 125.57 & 123.63 & 45.45 & 64.56 & 90.23 & 29.246 \\ 
\hline 
\hline \\[-1.8ex] 
\multicolumn{7}{r}{$^{*}$p$<$0.05; $^{**}$p$<$0.01; $^{***}$p$<$0.001} \\ 
\end{tabular} 
\endgroup 

\floatfoot{\textit{Notes:} The mental health scale sums five questions from the SF-36 and transforms the scale to range from 0-100. Individual level cluster robust standard errors are reported in the parentheses. The F-statistic null hypothesis is included financial hardship variables are jointly zero. R-Squared values are calculated using within variation only. Control variables used are the natural logarithm of equivalised household disposable income, employment status (full-time, part-time, other), a categorical variable for the neighbourhood quintile of disadvantage of residence, and the first lag of the PCS score. Disposable income (in constant 2019 dollars) is at the household level and is equivalised based on the number of members following OECD equivalence weights.}
\label{tab.mcs.fe.baseline.sf36mh}
\end{table}

	\begin{figure}[!hp]
\centering
\caption{Matched DID Results - Mental Health Scale Dependent Variable}
\includegraphics[scale=.6]{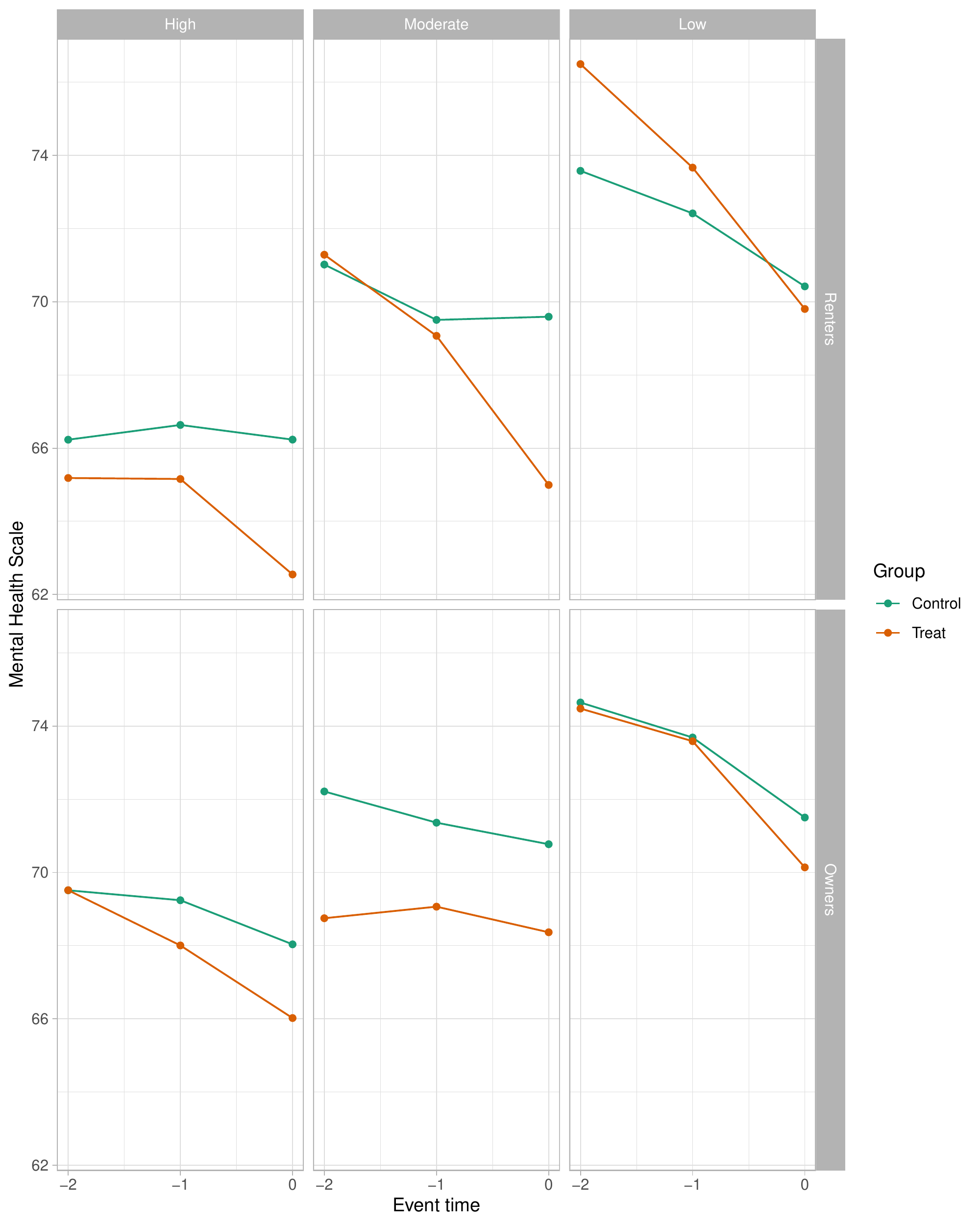}
\label{fig.results.match.trends.plot.rent.sf36mh}
\floatfoot{\textit{Notes:} Dependent variable is the mental health scale, rather than the MCS. Mental health scale predicted values plotted are conditional on event time and treatment group, and at the mean of control variables for the given time period. The pre-treatment periods are $t=-2$ and $t=-1$. In the pre-treatment periods individuals are in non-housing related financial hardship, but no housing related financial hardship. The post-treatment period is $t=0$. The treatment group has housing related financial hardship and non-housing related financial hardship at $t=0$, whereas the control group has non-housing related financial hardship only.}
\end{figure}

\begin{table}[!bh]
\centering
\caption{DID Hypothesis Testing - Alternative dependent variable}
\begingroup\normalsize
\begin{tabular}{llrrrrrr}
  \hline
Test &  & Renters:Low &  Mod &  High & Owners:Low & Mod & High \\ 
  \hline
\hline
Parallel Trends & Estimate & -1.67 & -0.70 & -0.43 & 0.06 & 1.17 & -1.24 \\ 
   & p-value & 0.26 & 0.69 & 0.75 & 0.95 & 0.57 & 0.42 \\ 
   \hline
Treatment Effects & Estimate & -3.53 & -4.86 & -2.64 & -1.20 & 1.06 & -2.02 \\ 
   & p-value & 0.05 & 0.01 & 0.06 & 0.36 & 0.62 & 0.18 \\ 
   \hline
\hline
\end{tabular}
\endgroup

\floatfoot{\textit{Notes:} Dependent variable is the mental health scale, rather than the MCS. All tests are performed with t-tests of equation (\ref{eq.did.estimation}) using individual level cluster robust standard errors. Parallel trends is tested with a null of $\gamma_{2}^{(h)} = 0$. Treatment effects is tested with a null of $\tau^{(h)} = 0$.}
\label{tab.het.results.did.sum.sf36mh}
\end{table}

\end{document}